\documentclass[aps,prd,twocolumn,showpacs,notitlepage,eqsecnum,
superscriptaddress,nofootinbib]{revtex4-1}

\usepackage[centertags]{amsmath} \usepackage{amssymb} \usepackage{latexsym}
\usepackage{enumerate} \usepackage{graphicx} \usepackage{mathrsfs}
\usepackage[colorlinks]{hyperref} \usepackage{stmaryrd} \usepackage{import}
\usepackage{tensor} \usepackage[usenames,dvipsnames]{xcolor} \usepackage{bm}
\usepackage{multirow} \usepackage{breakurl} \usepackage{float}
\usepackage{verbatim} \usepackage{mathrsfs}

\allowdisplaybreaks[1]

\definecolor{CiteColor}{rgb}{0,0.5,0} \hypersetup{citecolor=CiteColor}
\definecolor{RefColor}{rgb}{0.55,0,0} \hypersetup{linkcolor=RefColor}
\definecolor{darkgreen}{rgb}{0.2,0.7,0.2}

\begin{document}

\title{Reissner-Nordström perturbation framework with gravitational wave applications}

\author{Justin Y. J. Burton}
\affiliation{Department of Physics, 
Emory University, Atlanta, Georgia 30322, USA}
\affiliation{Department of Physics, 
Oxford College, Oxford, Georgia 30054, USA}
\affiliation{Department of Physics and Astronomy, 
State University of New York at Geneseo, New York 14454, USA}
\author{Thomas Osburn}
\email{tosburn@geneseo.edu}
\affiliation{Department of Physics and Astronomy, 
State University of New York at Geneseo, New York 14454, USA}
\affiliation{Department of Physics and Astronomy, 
University of North Carolina, Chapel Hill, North Carolina 27599, USA}
\affiliation{Department of Physics, 
Oxford College, Oxford, Georgia 30054, USA}

\begin{abstract}
We present a new convenient framework for modeling Reissner-Nordström black hole perturbations from charged distributions of matter. Using this framework, we quantify how gravitational wave observations of compact binary systems would be affected if one or both components were charged. Our approach streamlines the (linearized) Einstein-Maxwell equations through convenient master functions that we designed to ameliorate certain disadvantages of prior strategies. By solving our improved master equations with a point source, we are able to quantify the rate of orbital energy dissipation via electromagnetic and gravitational radiation. Through adiabatic and quasicircular approximations, we apply our dissipative calculations to determine trajectories for intermediate and extreme mass-ratio inspirals. By comparing trajectories and waveforms with varied charges to those with neutral components, we explore the potential effect of electric charge on gravitational wave signals. We observe that the case of opposite charge-to-mass ratios has the most dramatic impact. Our findings are largely interpreted through the lens of the upcoming LISA mission.
\end{abstract}

\maketitle

\section{Introduction}
\label{sec:intro}

Relativistic systems involving black holes comprise a diverse array of subjects at the forefront of astrophysics. Various types of black holes are involved in a wide variety of dramatic astronomical investigations ranging from active galactic nuclei~\cite{Rees84} to gravitational wave discoveries~\cite{LIGO1,LIGO2,LIGO3,LIGO4,LIGO5,LIGO7,LIGO8}. Naturally, it is valuable to explore how the fundamental properties of black holes affect the dynamics of related astronomical events. General relativity predicts that individual black holes are exhaustively described by only three properties: mass, spin, and electric charge. Often, a black hole's electric charge is assumed to be negligible in realistic scenarios because the surrounding plasma is capable of neutralization~\cite{Nara05}. However, compelling models exist to explain how black holes might become charged in a steady-state configuration. Many of these charged black hole models involve a rotating black hole in the presence of a magnetic field~\cite{Wald74,Ruff77,Puns98}, while other models consider more exotic scenarios such as interactions with charged dark matter~\cite{RujuETC90,CardETC16}. In pursuit of a more comprehensive understanding of how the properties of black holes affect their environment, this work refines black hole perturbation theory techniques to develop a robust formalism for modeling relativistic events involving charged black holes.

Our purpose here does not include further assessment of how black holes might originally become charged. Rather, we apply general relativity to predict the behavior of systems involving black holes with preexisting charge. For simplicity we consider nonrotating black holes, which are described by Reissner-Nordstr$\ddot{\text{o}}$m (RN) spacetime. To model dynamical scenarios using black hole perturbation theory, an RN black hole is assumed to have the strongest influence over nearby gravitational and electromagnetic fields. Additional influences are then accounted for by introducing small gravitational and electromagnetic perturbations to quantify the effect of nearby astronomical objects (see Sec.~\ref{sec:pert}). Note that, unlike post-Newtonian theory, these techniques do not require a slow motion approximation so that even highly relativistic systems are suitable candidates for this scheme provided that the size of the perturbations remains small compared to that of the black hole background. Perturbations of RN spacetime have been studied in the past~\cite{Zeri74,Monc74a,Monc74b,Monc75}, but those analyses involve certain disadvantages such as the assumption of a source-free environment or restriction of calculations to the frequency-domain (see Sec.~\ref{sec:mastback} for a detailed discussion). Here we present a mature and practical ``master function'' formalism that mitigates disadvantages encountered in prior work. A partial presentation of this formalism was made in~\cite{ZhuOsbu18}, but that analysis was restricted to an external environment consisting solely of a neutral point mass. This work generalizes that formalism to allow for arbitrary distributions of charged matter by carefully introducing appropriate electromagnetic source terms into the mathematical definitions of the master functions (see Sec.~\ref{sec:master}).

To focus our analysis, we consider an example system in which a charged compact object orbits the RN black hole. This lower mass binary component could be a charged neutron star~\cite{Cohe75} or smaller black hole. At linear order in perturbation theory, such a compact object is represented accurately by a point particle. The rapid radial infall of point particles into an RN black hole has been studied previously~\cite{CardETC16,John73,John74} (as have similar collisions of charged black holes with comparable masses~\cite{Zilh12,Zilh14}), but relativistic charged binary encounters involving sustained orbital configurations have not been modeled before. Here we consider such a sustained encounter by calculating the gravitational and electromagnetic radiation emitted by a charged compact object on a circular orbit in RN spacetime for the first time. Compact binary systems like these are crucial sources for gravitational wave astronomy~\cite{LIGO1,LIGO2,LIGO3,LIGO4,LIGO5,LIGO7,LIGO8,LIGO6}. Therefore, as an application of the theoretical and computational techniques presented here, we perform the first calculations to quantify how electric charge on one or both binary components would affect gravitational wave observations. We consider intermediate and extreme mass-ratio binary systems in this analysis (comparable mass binaries are not well represented by linear perturbation theory). Extreme mass-ratio binaries are valuable sources for the upcoming Laser Interferometer Space Antenna (LISA) mission~\cite{LISA}, while intermediate mass-ratio binaries are important for both LISA and LIGO-Virgo~\cite{aLIGO,aVIRGO} observations depending on the total mass of the system~\cite{ZhuOsbu18}.

Compact binary systems such as these have orbits that decay over time because electromagnetic and gravitational waves described by the master functions in our formalism carry away orbital energy. This radiation reaction is formally quantified by considering how the point particle interacts with its perturbations through a mechanism called the self-force~\cite{MinoSasaTana97,QuinWald97,Pois11}. While preliminary work has been conducted regarding the self-force in RN spacetime~\cite{Bara00,Burk01,ZimmPois14,Linz14,Zimm15,CastVegaWard18}, we avoid the technical complexities associated with self-force calculations by making an adiabatic approximation~\cite{Poun05} that leverages the radiative energy decay rate to generate equations of motion governing the quasicircular inspiral of the charged compact object toward the RN black hole (see Sec.~\ref{sec:num}). After calculating the master functions through our perturbative analysis and determining the inspiral trajectory through adiabatic and quasicircular approximations, we finally generate waveforms describing the gravitational waves emitted by electrically charged compact binary systems for the first time. By comparing to the case with neutral binary components (which reflects existing models), we are able to quantify how critical it would be to account for electric charge in practical waveform templates if gravitational waves from charged binary systems were detected (see Sec.~\ref{sec:results}). 

\begin{widetext}

\section{Perturbations of charged black holes}
\label{sec:pert}

\subsection{Overview}

We model the environment surrounding a charged black hole (with mass $M$ and charge $Q$) by considering first-order perturbations of the Einstein-Maxwell equations in RN spacetime. In this scheme, a small parameter, $\epsilon$, is used to expand the spacetime metric, $g_{\alpha\beta}$, and the electromagnetic potential four-vector, $A_\alpha$,
\begin{align}
&g_{\alpha\beta} = g^{(0)}_{\alpha\beta} + g^{(1)}_{\alpha\beta} + \mathcal{O}(\epsilon^2) ,
\\
&A_\alpha = A^{(0)}_\alpha + A^{(1)}_\alpha + \mathcal{O}(\epsilon^2) .
\end{align}
$A_\alpha$ and $g_{\alpha\beta}$ are governed by the Einstein equations, 
\begin{align}
\label{eq:Ein}
G_{\alpha\beta} = 8\pi T_{\alpha\beta}  ,
\end{align}
and Maxwell's equations,
\begin{align}
\label{eq:Max}
\nabla_\beta F^{\alpha\beta} = 4\pi J^\alpha ,
\end{align}
where $F^{\alpha\beta}$ is the electromagnetic field tensor, 
\begin{align}
F_{\alpha\beta} = \nabla_\alpha A_\beta - \nabla_\beta A_\alpha, 
\end{align}
$J^\alpha$ is the current density four-vector, $G_{\alpha\beta}$ is the Einstein tensor, and $T_{\alpha\beta}$ is the stress-energy tensor, 
\begin{align}
&T^{\alpha\beta} = T^{\alpha\beta}_\text{matter} + \frac{1}{4\pi}\left( F^{\alpha}_{\;\;\gamma} F^{\beta\gamma} - \frac{1}{4}g^{\alpha\beta}F_{\gamma\nu}F^{\gamma\nu} \right).
\label{eq:Tab}
\end{align}
When the charged black hole's surroundings consist of a lower mass compact object (with mass $\mu$), the small expansion parameter is the mass-ratio: $\epsilon = \mu/M$. The leading terms in the expansion, $g^{(0)}_{\alpha\beta}$ and $A^{(0)}_\alpha$, are exact solutions of Eqs.~\eqref{eq:Ein} and \eqref{eq:Max} describing the central black hole. Electric charge is incorporated by adopting the RN metric for $g^{(0)}_{\alpha\beta}$ (we utilize Boyer-Lindquist coordinates, $x^\alpha$),
\begin{align}
g^{(0)}_{\alpha\beta}\, dx^\alpha dx^\beta &= -f \, dt^2 + \frac{1}{f} \,dr^2 + r^2\left(d\theta^2 + \sin^2\theta \, d\varphi^2 \right) ,
\\
f  &\equiv 1-\frac{2M}{r}+\frac{Q^2}{r^2} .
\end{align}
$A^{(0)}_\alpha$ is a vacuum solution of Maxwell's equations compatible with the RN metric,
\begin{align}
A^{(0)}_t &= -\frac{Q}{r}, 
\\
A^{(0)}_r &= A^{(0)}_\theta = A^{(0)}_\varphi = 0 . \notag
\end{align}

The first-order gravitational and electromagnetic perturbations, $g^{(1)}_{\alpha\beta}$ and $A^{(1)}_\alpha$, are determined by expanding Eqs.~\eqref{eq:Ein} and \eqref{eq:Max} through linear order in $\epsilon$. Rather than explicitly presenting the perturbed field equations here in covariant form (see \cite{Bini07}), we first perform multipole decompositions of $g^{(1)}_{\alpha\beta}$ and $A^{(1)}_\alpha$ and then use separation of variables to present the differential equations governing each mode. At first order in $\epsilon$, a smaller binary component can be represented as a charged point mass with position $x_p^\alpha$, four-velocity $u^\alpha$, mass $\mu$, and charge $q$. For such a point particle, the sources driving Eqs. \eqref{eq:Ein} and \eqref{eq:Max} are described by Dirac delta functions, 
\begin{align}
\label{eq:J}
& J^\alpha = q \frac{u^\alpha}{u^t r_p^2 \sin{\theta_p}}\, \delta(r-r_p) \, \delta(\theta-\theta_p) \, \delta(\varphi-\varphi_p) ,
\\
\label{eq:Tmat}
&T^{\alpha\beta}_\text{matter} = \mu \frac{u^\alpha u^\beta}{u^t r_p^2 \sin{\theta_p}} \, \delta(r-r_p) \, \delta(\theta-\theta_p) \, \delta(\varphi-\varphi_p) .
\end{align}
We later narrow our focus even further by restricting the point particle's motion to a circular equatorial orbit ($r_p = \text{constant}$, $\theta_p = \pi/2$, $\varphi_p = \Omega \, t$). Note, however, that this formalism is not in any way restricted to the point particle example studied here. 

\subsection{Multipole decompositions}

\label{sec:multipole}

The electromagnetic perturbation, $A^{(1)}_\alpha$, and current density, $J^\alpha$, are decomposed into vector spherical harmonics,
\begin{align}
\label{eq:atr}
&A^{(1)}_b(t,r,\theta,\varphi) = \sum_{lm} a^{lm}_b(r) \,Y^{lm}(\theta,\varphi) \, e^{-i \omega t} ,
\\
\label{eq:athph}
&A^{(1)}_B(t,r,\theta,\varphi) = \sum_{lm} \left[a^{lm}_\sharp(r) \, Y_B^{lm}(\theta,\varphi) + a_{lm}^\text{odd}(r) \, X_B^{lm}(\theta,\varphi)\right] e^{-i \omega t} ,
\\
&J^{a}(t,r,\theta,\varphi) = \frac{1}{4\pi} \sum_{lm} \mathcal{J}^a_{lm}(r)\, Y^{lm}(\theta,\varphi) \,e^{-i\omega t} ,
\\
&J^{A}(t,r,\theta,\varphi) = \frac{1}{4\pi r^2} \sum_{lm} \left[\mathcal{J}^{\sharp}_{lm}(r)\, Y^A_{lm}(\theta,\varphi) + \mathcal{J}^{\flat}_{lm}(r)\, X^A_{lm}(\theta,\varphi) \right] e^{-i\omega t} .
\end{align}
We have adopted the notation of Martel and Poisson \cite{MartPois05} where lowercase Latin indices ($a$,$b$) refer to $t$ and $r$ tensor components and uppercase Latin indices ($A$,$B$) refer to $\theta$ and $\varphi$ tensor components. Similarly, the metric perturbation, $g^{(1)}_{\alpha\beta}$, and material stress-energy tensor, $T^{\alpha\beta}_\text{matter}$, are decomposed into tensor spherical harmonics, 
\begin{align}
\label{eq:httrr}
&g^{(1)}_{ab}(t,r,\theta,\varphi) = \sum_{lm} h^{lm}_{ab}(r) \,Y^{lm}(\theta,\varphi) \, e^{-i \omega t} ,
\\
\label{eq:htr}
&g^{(1)}_{aB}(t,r,\theta,\varphi) = \sum_{lm} \left[ j^{lm}_{a}(r) \,Y_B^{lm}(\theta,\varphi) +  h^{lm}_{a}(r) \,X_B^{lm}(\theta,\varphi) \right] e^{-i \omega t} ,
\\
\label{eq:hthph}
&g^{(1)}_{AB}(t,r,\theta,\varphi) = \sum_{lm} \left[ r^2 K^{lm}(r)\, \Omega_{AB}(\theta,\varphi)\, Y^{lm}(\theta,\varphi) + r^2 G^{lm}(r) \, Y^{lm}_{AB}(\theta,\varphi) +  h_2^{lm}(r) \,X_{AB}^{lm}(\theta,\varphi) \right] e^{-i \omega t} ,
\\
&T^{ab}_\text{matter}(t,r,\theta,\varphi) = \frac{1}{8\pi} \sum_{lm} \mathcal{Q}^{ab}_{lm}(r) \, Y^{lm}(\theta,\varphi) \,e^{-i\omega t} ,
\\
&T^{aB}_\text{matter}(t,r,\theta,\varphi) = \frac{1}{16\pi r^2} \sum_{lm} \left[\mathcal{Q}^{a}_{lm}(r)\, Y^B_{lm}(\theta,\varphi) + \mathcal{P}^{a}_{lm}(r)\, X^B_{lm}(\theta,\varphi) \right] e^{-i\omega t} ,
\\
\label{eq:TAB}
&T^{AB}_\text{matter}(t,r,\theta,\varphi) = \frac{1}{16\pi r^2} \sum_{lm} \left[\mathcal{Q}^{\flat}_{lm}(r)\, \Omega^{AB}(\theta,\varphi) \, Y^{lm}(\theta,\varphi) + \frac{1}{r^2}\mathcal{Q}^{\sharp}_{lm}(r)\, Y^{AB}_{lm}(\theta,\varphi) + \frac{2}{r^2}\mathcal{P}_{lm}(r)\, X^{AB}_{lm}(\theta,\varphi) \right] e^{-i\omega t} .
\end{align}
The definitions of all angular functions appearing in Eqs.~\eqref{eq:atr}-\eqref{eq:TAB} are provided in \cite{MartPois05}. We have also entered the frequency-domain by assuming (for brevity) sinusoidal time dependence (this simple time representation is easily generalizable via the Fourier transform). However, it is important to note that the formalism presented here does not require frequency-domain calculations; rather, the frequency-domain is a convenient option for our circular motion example. For arbitrary sources, time-domain generalizability ($-i\omega \rightarrow \frac{\partial}{\partial t}$) is a major advantage afforded by the methods presented here (see Sec.~\ref{sec:master}). Expressions for the multipole modes of the sources ($\mathcal{J}^a_{lm}$, $\mathcal{J}^{\sharp}_{lm}$, $\mathcal{J}^{\flat}_{lm}$, $\mathcal{Q}^{ab}_{lm}$, $\mathcal{Q}^{a}_{lm}$, $\mathcal{P}^{a}_{lm}$, $\mathcal{Q}^{\flat}_{lm}$, $\mathcal{Q}^{\sharp}_{lm}$, and $\mathcal{P}_{lm}$) are given in Appendix \ref{sec:sources}.

Maxwell's equations are enforced by expanding Eq.~\eqref{eq:Max} through the first power of $\epsilon$ for each multipole ($l$,$m$) mode, 
\begin{align}
\label{eq:odd1}
\mathcal{J}^{\flat}_{lm} &= f\frac{d^2 a_{lm}^\text{odd}}{dr^2} +\frac{2(Mr-Q^2)}{r^3}\frac{d a_{lm}^\text{odd}}{dr} + \left( \frac{\omega^2}{f} -\frac{l(l+1)}{r^2} \right) a_{lm}^\text{odd} + \frac{Q}{r^2}\frac{d h_t^{lm}}{dr} - \frac{2Q}{r^3} h_t^{lm} + \frac{i\omega Q}{r^2} h_r^{lm} ,
\\
\label{eq:even1}
\mathcal{J}^t_{lm} &= \frac{d^2 a_t^{lm}}{dr^2} + \frac{2}{r}\frac{d a_t^{lm}}{dr} + i\omega \frac{da_r^{lm}}{dr} -\frac{l(l+1)}{r^2f} a_t^{lm} + \frac{2i\omega }{r} a_r^{lm}  \notag
\\&\qquad\qquad\qquad\qquad +\frac{Q}{2r^2f}\frac{d h_{tt}^{lm}}{dr} - \frac{fQ}{2r^2} \frac{d h_{rr}^{lm}}{dr} + \frac{Q}{r^2} \frac{dK^{lm}}{dr} +\frac{Q(Q^2-M r)}{r^5 f^2}  \left( h_{tt}^{lm} + f^2 h_{rr}^{lm} \right) ,
\\
\label{eq:even2}
\mathcal{J}^r_{lm} &= i \omega \frac{da_t^{lm}}{dr} + \frac{i\omega Q}{2 r^2f} h_{tt}^{lm} - \frac{i\omega f Q}{2 r^2} h_{rr}^{lm} + \frac{i\omega Q}{r^2} K^{lm} - \left( \omega^2-\frac{f}{r^2}l(l+1) \right) a_r^{lm} ,
\\
\label{eq:even3}
\mathcal{J}^{\sharp}_{lm} &= f \frac{da_r^{lm}}{dr} + \frac{i\omega}{f} a_t^{lm} - \frac{2(Q^2-Mr)}{r^3} a_r^{lm} .
\end{align}
Similarly, Einstein's equations are enforced by expanding Eq.~\eqref{eq:Ein} through the first power of $\epsilon$ for each mode, 
\begin{align}
&\mathcal{P}^t_{lm} = \frac{d^2h_t^{lm}}{dr^2} + i\omega \frac{dh_r^{lm}}{dr} +\frac{2i\omega}{r} h_r^{lm} - \frac{l(l+1)r^2-4 M r+2 Q^2}{f r^4} h_t^{lm} + \frac{4Q}{r^2}\frac{da^\text{odd}_{lm}}{dr} ,
\label{eq:odd2}
\\
&\mathcal{P}^r_{lm} = i\omega \frac{dh_t^{lm}}{dr}-\frac{2 i \omega}{r} h_t^{lm} - \left(\omega^2 -\frac{f}{r^2}(l+2)(l-1) \right) h_r^{lm} + \frac{4i\omega Q}{r^2} a^\text{odd}_{lm} ,
\label{eq:odd3}
\\
&\mathcal{P}_{lm} = f\frac{dh_r^{lm}}{dr} - \frac{2(Q^2-M r)}{r^3} h_r^{lm} + \frac{i\omega}{f}h_t^{lm},
\label{eq:odd4}
\\
\label{eq:even4}
&\mathcal{Q}^{tt}_{lm} = -\frac{d^2K^{lm}}{dr^2} - \frac{2Q^2+r(3r-5M)}{fr^3}\frac{dK^{lm}}{dr}+\frac{f}{r} \frac{dh_{rr}^{lm}}{dr}+\frac{(l+2)(l-1)}{2fr^2} K^{lm} \notag
\\&\qquad\qquad\qquad\qquad - \frac{4Q^2-r(4M+r(l(l+1)+2))}{2 r^4} h_{rr}^{lm} -\frac{Q^2}{r^4f^2} h_{tt}^{lm} -\frac{2Q}{r^2 f}\frac{da_{t}^{lm}}{dr} - \frac{2i\omega Q}{r^2f} a_r^{lm} ,
\\
&\mathcal{Q}^{tr}_{lm} = -i\omega \frac{dK^{lm}}{dr}-\frac{l(l+1)}{2r^2} h^{lm}_{tr} +\frac{i\omega f}{r} h_{rr}^{lm} -\frac{i\omega (2Q^2+r(r-3M))}{r^3f} K^{lm} ,
\\
&\mathcal{Q}^{rr}_{lm} = \frac{f(r-M)}{r^2}\frac{dK^{lm}}{dr} - \frac{f}{r}\frac{dh_{tt}^{lm}}{dr}+\left(\omega^2 - \frac{f(l+2)(l-1)}{2r^2}\right)K^{lm} -\frac{f^2}{r^2}h_{rr}^{lm}  \notag
\\&\qquad\qquad\qquad\qquad - \frac{2Q^2-r(4M+rl(l+1))}{2r^4} h_{tt}^{lm} - \frac{2i\omega f}{r} h_{tr}^{lm} +\frac{2fQ}{r^2}\frac{da_t^{lm}}{dr} + \frac{2i\omega fQ}{r^2}a_r^{lm} ,
\label{eq:even6}
\\
&\mathcal{Q}^t_{lm} = -\frac{d h_{tr}^{lm}}{dr}-i \omega h_{rr}^{lm} + \frac{2(Q^2-M r)}{r^3f}h_{tr}^{lm} -\frac{i\omega}{f} K^{lm} - \frac{4 Q}{r^2}a_r^{lm} ,
\label{eq:even7}
\\
&\mathcal{Q}^r_{lm} = \frac{d h_{tt}^{lm}}{dr} -f \frac{dK^{lm}}{dr} - \frac{r-M}{r^2f} h_{tt}^{lm} + i\omega h_{tr} +\frac{f(r-M)}{r^2} h_{rr}^{lm} + \frac{4 Q}{r^2} a_t^{lm} ,
\label{eq:even8}
\\
&\mathcal{Q}^\flat_{lm} =  f \frac{d^2K^{lm}}{dr^2} - \frac{d^2h_{tt}^{lm}}{dr^2} +\frac{2(r-M)}{r^2}\frac{dK^{lm}}{dr} -\left( \frac{2}{r} -\frac{r-M}{r^2f}\right)\frac{dh_{tt}^{lm}}{dr}-\frac{f(r-M)}{r^2}\frac{dh_{rr}^{lm}}{dr} \notag
\\&\qquad\qquad\qquad\qquad -2i\omega \frac{dh_{tr}^{lm}}{dr} +\frac{l(l+1)}{2r^2f}\left( h_{tt}^{lm} -f^2 h_{rr}^{lm} \right)+ \frac{2(r-M)(Q^2-Mr)}{f^2 r^5}\left( h_{tt}^{lm} +f^2 h_{rr}^{lm} \right) \notag
\\&\qquad\qquad\qquad\qquad\qquad\qquad\qquad  +\omega^2 h_{rr}^{lm} +\frac{\omega^2}{f}K^{lm} -\frac{2i\omega(r-M)}{r^2f}h_{tr}^{lm}-\frac{4Q}{r^2}\frac{da_t^{lm}}{dr}-\frac{4i\omega Q}{r^2} a_r^{lm} ,
\label{eq:even9}
\\
&\mathcal{Q}^\sharp_{lm} = \frac{1}{f} h_{tt}^{lm} - f h_{rr}^{lm} .
\label{eq:even10}
\end{align}
The metric perturbation is not unique due to the 4 degrees of gauge freedom. We adopt the Regge-Wheeler gauge where $h^{lm}_2=j^{lm}_t=j^{lm}_r=G^{lm}=0$. Similarly, there is 1 degree of electromagnetic gauge freedom, and we adopt a gauge where $a^{lm}_\sharp=0$. Through this description, the fundamental problem reduces to solving (for each mode) Eqs.~\eqref{eq:odd1}-\eqref{eq:even10} for the 9 nonvanishing radial functions: $a^{lm}_t$, $a^{lm}_r$, $a_{lm}^\text{odd}$,  $h^{lm}_{tt}$, $h^{lm}_{tr}$, $h^{lm}_{rr}$, $K^{lm}$, $h^{lm}_{t}$, and $h^{lm}_{r}$.
Then the overall solution is constructed through Eqs.~\eqref{eq:atr}-\eqref{eq:athph} and \eqref{eq:httrr}-\eqref{eq:hthph}. 

The source terms are required to obey certain constraints implied by physical conservation laws. One such constraint is conservation of charge, $\nabla_\alpha J^\alpha = 0$, which reduces to a relationship between the modes of the current density,
\begin{align}
\label{eq:consQ}
0 &= i\omega \mathcal{J}^t_{lm} - \frac{2}{r} \mathcal{J}^r_{lm} + \frac{l(l+1)}{r^2}\mathcal{J}^\sharp_{lm} - \frac{d \mathcal{J}^r_{lm}}{dr} .
\end{align}
As Maxwell's equations themselves imply conservation of charge, Eq.~\eqref{eq:consQ} can also be derived by direct substitution of the field equations. Similarly, there is conservation of stress-energy, $\nabla_\alpha T^{\alpha\beta} = 0$, which also reduces to relationships between the source modes
\begin{align}
\label{eq:consSE1}
0 &= i\omega \mathcal{P}^t_{lm} - \frac{2}{r} \mathcal{P}^r_{lm}+\frac{(l+2)(l-1)}{r^2} \mathcal{P}_{lm} - \frac{d\mathcal{P}^r_{lm}}{dr} ,
\\
\label{eq:consSE2}
0 &= \frac{2Q}{r^2 f}\mathcal{J}^r_{lm} + i\omega \mathcal{Q}^{tt}_{lm}+\frac{2(M-r)}{r^2 f} \mathcal{Q}^{tr}_{lm}+\frac{l(l+1)}{2r^2}\mathcal{Q}^{t}_{lm} - \frac{d\mathcal{Q}^{tr}}{dr} ,
\\
\label{eq:consSE3}
0 &= \frac{2Qf}{r^2}\mathcal{J}^t_{lm} + \frac{f(Q^2-Mr)}{r^3} \mathcal{Q}^{tt}_{lm}+i\omega \mathcal{Q}^{tr}_{lm}+\left( \frac{r-M}{r^2 f}-\frac{3}{r} \right) \mathcal{Q}^{rr}_{lm}+\frac{l(l+1)}{2r^2} \mathcal{Q}^{r}_{lm} +\frac{f}{r} \mathcal{Q}^\flat_{lm} - \frac{d\mathcal{Q}^{rr}_{lm}}{dr} ,
\\
\label{eq:consSE4}
0 &= i\omega \mathcal{Q}^{t}_{lm} - \frac{2}{r} \mathcal{Q}^r_{lm}-\mathcal{Q}^\flat_{lm}+\frac{(l+2)(l-1)}{2r^2} \mathcal{Q}^\sharp_{lm}-\frac{d\mathcal{Q}^r_{lm}}{dr} .
\end{align}
As the Einstein equations themselves imply conservation of stress-energy, Eqs.~\eqref{eq:consSE1}-\eqref{eq:consSE4} can also be derived by direct substitution of the field equations. It is perhaps surprising that current density modes, like $\mathcal{J}^t_{lm}$ and $\mathcal{J}^r_{lm}$, appear at all in Eqs.~\eqref{eq:consSE1}-\eqref{eq:consSE4} considering that the current density is not explicitly related to the stress-energy tensor. Direct consideration of stress-energy conservation necessarily involves the electromagnetic fields that appear in the stress-energy tensor. However, explicit appearance of any fields in the constraints can be eliminated through careful substitution of Maxwell’s equations. Therefore, Eqs.~\eqref{eq:consSE1}-\eqref{eq:consSE4} involve a combination of stress-energy conservation and Maxwell’s equations. Point particle source terms are guaranteed to satisfy Eqs.~\eqref{eq:consQ}-\eqref{eq:consSE4} when orbital motion is determined through the Lorentz force law.

\section{Master function formalism}
\label{sec:master}

\subsection{Background}
\label{sec:mastback}

Solving for the 9 nonvanishing radial functions presents a challenge because they are governed by 14 field equations that are not entirely independent. One welcome simplification recognizes that Eqs.~\eqref{eq:odd1} and \eqref{eq:odd2}-\eqref{eq:odd4} are not at all coupled to Eqs.~\eqref{eq:even1}-\eqref{eq:even3} and \eqref{eq:even4}-\eqref{eq:even10} (and vice versa). These two separate sets of field equations are referred to as the ``odd-parity'' system of 4 coupled equations (Eqs.~\eqref{eq:odd1} and \eqref{eq:odd2}-\eqref{eq:odd4}) and the ``even-parity'' system of 10 coupled equations (Eqs.~\eqref{eq:even1}-\eqref{eq:even3} and \eqref{eq:even4}-\eqref{eq:even10}), which are named based on how their respective angular functions behave under a parity transformation. Each parity is then handled separately when solving the field equations for a specific spherical harmonic mode. The even- and odd-parity radial functions are later reunited when summing over multipoles. For sources confined to the equatorial plane (which includes the point particle in circular motion example presented here), the odd-parity source terms vanish when $l+m$ is an even number, and the even-parity source terms vanish when $l+m$ is an odd number. Therefore, it is natural in these scenarios to consider only the odd-parity field equations when $l+m$ is odd and only the even-parity field equations when $l+m$ is even.

In addition to the odd- and even-parity split, we seek additional simplifications through which the numerous gravitational and electromagnetic fields (for a given parity) are constructed from a decreased number of streamlined elements called ``master functions.'' These ``master functions'' are governed by ``master equations'' whose details are motivated by the fundamental field equations. RN perturbations have been described using master functions previously by Moncrief in the homogeneous case~\cite{Monc74a,Monc74b,Monc75} and by Zerilli~\cite{Zeri74} in the nonhomogeneous case. Zerilli's formalism has the advantage that it includes source terms, but it is not well suited for time-domain calculations because frequency ($\omega$) appears in the denominator of his master equations (and his metric reconstruction equations). Also, Zerilli's master equations have consequentially (and perhaps inconveniently) different forms in the odd- and even-parity cases. Conversely, Moncrief's formalism has the advantages that it is well suited for time-domain calculations (frequency does not appear in the denominator) and that his even- and odd-parity master equations share a convenient form. Unfortunately, Moncrief did not consider source terms in his formalism. Here we present a novel alternative that combines the advantages of Zerilli's and Moncrief's formalisms while avoiding their disadvantages. Our formalism is equivalent to generalizing Moncrief's master equations to the nonhomogeneous case. A partial presentation of this formalism was made in~\cite{ZhuOsbu18}, but that work did not consider electromagnetic source terms. This section completes that presentation by revealing how any nonvanishing electromagnetic sources are necessarily incorporated into the master function definitions and master equations. The low multipole modes ($l=1$ and $l=0$) modes require special treatment; see Appendix~\ref{sec:low-multipole} for more details.

\subsection{Odd-parity master equations}

For each mode, the odd-parity perturbations are constructed from two master functions. The gravitational master function, $h_{lm}^\text{odd}$, determines $h_t^{lm}$ and $h_r^{lm}$,
\begin{align}
\label{eq:ht}
h_t^{lm} &= \frac{rf}{2}\frac{dh^\text{odd}_{lm}}{dr} + \frac{f}{2} h^\text{odd}_{lm} - \frac{f r^2}{(l+2)(l-1)} \mathcal{P}^t_{lm} ,
\\
\label{eq:hr}
h_r^{lm} &= -\frac{i\omega r}{2f} h^\text{odd}_{lm} + \frac{r^2}{f(l+2)(l-1)} \mathcal{P}^r_{lm} .
\end{align}
The electromagnetic master function, $a^\text{odd}_{lm}$, has already appeared in the multipole decomposition. These master functions are governed by master equations that follow from substituting Eqs.~\eqref{eq:ht} and \eqref{eq:hr} into Eqs.~\eqref{eq:odd1} and \eqref{eq:odd3},
\begin{align}
& \frac{4 f (Mr-Q^2)}{r^2(l+2)(l-1)} \mathcal{P}^t_{lm} - \frac{2 i\omega r}{(l+2)(l-1)} \mathcal{P}^r_{lm} + \frac{2 f^2 r}{(l+2)(l-1)} \frac{d\mathcal{P}^t_{lm}}{dr} = \notag
\\&\qquad\qquad\qquad\qquad\qquad\qquad\qquad\qquad \frac{d^2 h^\text{odd}_{lm}}{dr_*^2} + \left( \omega^2 - \frac{f(l(l+1)r^2-6Mr+4Q^2)}{r^4} \right) h^\text{odd}_{lm} + \frac{8 f Q}{r^3}a^\text{odd}_{lm} ,
\label{eq:hodd}
\\
\label{eq:aodd}
& \qquad\qquad\qquad f \mathcal{J}^{\flat}_{lm} = \frac{d^2 a^\text{odd}_{lm}}{dr_*^2} + \left( \omega^2 - \frac{f(l(l+1)r^2+4Q^2)}{r^4} \right) a^\text{odd}_{lm} + \frac{fQ(l+2)(l-1)}{2 r^3} h^\text{odd}_{lm} .
\end{align}
Here $r_*$ is the tortoise coordinate, which is related to $r$ through the differential equation $\dfrac{dr_*}{dr} = f^{-1}$ with the following solution:
\begin{align}
r_* &= r+ \frac{r_+^2}{r_+-r_-}\ln\left( \frac{r-r_+}{M} \right)- \frac{r_-^2}{r_+-r_-}\ln\left( \frac{r-r_-}{M} \right) .
\end{align}
The constants $r_\pm$ are the radii of the RN black hole's inner ($-$) and outer ($+$) event horizons,
\begin{align}
r_\pm = M \pm \sqrt{M^2-Q^2} .
\end{align}
We find it convenient to write the master equations (Eqs.~\eqref{eq:hodd} and \eqref{eq:aodd}) in matrix form, 
\begin{align}
\label{eq:masterOdd}
\left( \frac{d^2}{dr_*^2} + \omega^2 + \left[\begin{array}{cc} \alpha_l^\text{odd} & \beta_l^\text{odd} \\ \gamma_l^\text{odd} & \sigma_l^\text{odd} \end{array} \right] \right) \left[\begin{array}{c} h_{lm}^\text{odd} \\ a_{lm}^\text{odd} \end{array} \right] &= \left[\begin{array}{c} S_{lm}^\text{odd} \\ Z_{lm}^\text{odd} \end{array} \right] ,
\end{align}
where the coefficients and sources are determined through comparison of Eq.~\eqref{eq:masterOdd} with Eqs.~\eqref{eq:hodd} and \eqref{eq:aodd},
\begin{align}
&\alpha_l^\text{odd}(r) = -\frac{f(l(l+1)r^2-6Mr+4Q^2)}{r^4} , 
\\
&\beta_l^\text{odd}(r) = \frac{8 f Q}{r^3} ,
\\
&\sigma_l^\text{odd}(r) = - \frac{f(l(l+1)r^2+4Q^2)}{r^4} , 
\\
&\gamma_l^\text{odd}(r) = \frac{fQ(l+2)(l-1)}{2 r^3} ,
\\
&S_{lm}^\text{odd}(r) = \frac{4 f (Mr-Q^2)}{r^2(l+2)(l-1)} \mathcal{P}^t_{lm} - \frac{2 i\omega r}{(l+2)(l-1)} \mathcal{P}^r_{lm} + \frac{2 f^2 r}{(l+2)(l-1)} \frac{d\mathcal{P}^t_{lm}}{dr}, 
\\
&Z_{lm}^\text{odd}(r) = f \mathcal{J}^{\flat}_{lm} .
\end{align}
The forms of Eqs.~\eqref{eq:ht} and \eqref{eq:hr} were carefully selected so that, although Eqs.~\eqref{eq:odd2} and \eqref{eq:odd4} do not enter the master equations directly (unlike Eqs.~\eqref{eq:odd1} and \eqref{eq:odd3}),  all four odd-parity field equations are automatically satisfied when Eq.~\eqref{eq:masterOdd} is solved for $ h_{lm}^\text{odd}$ and $a_{lm}^\text{odd}$.

\subsection{Even-parity master equations}

Despite their greater complexity, the even-parity master equations follow from similar logic. The first step in relating the even-parity fields to the even-parity master functions is to express $h^{lm}_{tt}$, $h^{lm}_{tr}$, and $h^{lm}_{rr}$ in terms of $K^{lm}$,
\begin{align}
\label{eq:hrr}
&h_{rr}^{lm} = \frac{r^2}{2f}\frac{d^2K^{lm}}{dr^2}+\frac{r-M}{f^2}\frac{dK^{lm}}{dr}+\left( \frac{r^2 \omega^2}{2f^3} -\frac{(l+2)(l-1)}{2f^2} \right)K^{lm} -\frac{2Q}{f^2}\frac{da_t^{lm}}{dr} -\frac{4Q}{rf^2}a_t^{lm} \notag
\\&\qquad\qquad - \frac{2i\omega Q}{f^2} a_r^{lm}+\frac{r}{2f}\frac{d\mathcal{Q}^\sharp_{lm}}{dr}+\frac{r^2}{2f} \mathcal{Q}^{tt}_{lm} - \frac{4Q^2-r(12M+r(l(l+1)-4))}{4r^2f^2} \mathcal{Q}^\sharp_{lm} -\frac{r^2}{2f^3}\mathcal{Q}^{rr}_{lm} - \frac{r}{f^2} \mathcal{Q}^r_{lm} ,
\\
\label{eq:htr}
&h_{tr}^{lm} = \frac{2}{l(l+1)}\left( i\omega r f h_{rr}^{lm} - i\omega r^2\frac{dK^{lm}}{dr} - \frac{i\omega(2Q^2+r(r-3M))}{r f} K^{lm} -r^2 Q^{tr}_{lm} \right) ,
\\
\label{eq:htt}
&h_{tt}^{lm} = f^2 h_{rr}^{lm} +f Q^\sharp_{lm} .
\end{align}
It can be shown that Eqs.~\eqref{eq:hrr}-\eqref{eq:htt} follow from linear combinations of Eqs.~\eqref{eq:even4}-\eqref{eq:even10} and their $r$ derivatives. Then the even-parity gravitational master function, $h_{lm}^\text{even}$, enters through a relationship with $K^{lm}$,
\begin{align}
\label{eq:heven}
&K^{lm} = \left[ 2 r \left(-2 Q^2 r (2 M+\lambda  r)+(\lambda +1) r^3 (3 M+\lambda  r)+2 Q^4\right) \right]^{-1} \bigg( 2 f (\lambda +1) r^3 \left(r (3 M+\lambda  r)-2 Q^2\right) \frac{dh^\text{even}_{lm}}{dr} \notag
\\&\qquad\qquad\qquad -(\lambda + 1) \left[-2 r^2 \left(6 M^2+3 \lambda  M r+\lambda  (\lambda +1) r^2\right)+2 Q^2 r (11 M+2 (\lambda -1) r)-8 Q^4\right] h^\text{even}_{lm} \notag
\\&\qquad\qquad\qquad\qquad\qquad\qquad -4 r^5 f Q \frac{da_t^{lm}}{dr} - 4 i \omega r^5 f Q a_r^{lm} -2 r^7 f^2 \mathcal{Q}^{tt}_{lm} -2 r^3 f Q^2 \mathcal{Q}^\sharp_{lm} \bigg) ,
\end{align}
where $\lambda\equiv (l+2)(l-1)/2$. Similarly, $a_t^{lm}$ and $a_r^{lm}$ are expressed in terms of the electromagnetic master function $a^\text{even}_{lm}$,
\begin{align}
\label{eq:at}
& a_t^{lm} =  f \frac{d}{dr}\left(a^\text{even}_{lm} +\frac{Q}{2 r} h^\text{even}_{lm} \right) - \frac{r^2 f}{2(\lambda+1)}\mathcal{J}^t_{lm} ,
\\
\label{eq:ar}
&a_r^{lm} = -\frac{i \omega}{f} \left( a^\text{even}_{lm} +\frac{Q}{2 r} h^\text{even}_{lm} \right) + \frac{r^2}{2f(\lambda+1)} \mathcal{J}^r_{lm} .
\end{align}
The forms of Eqs.~\eqref{eq:heven}-\eqref{eq:ar} were carefully selected so that the 6 nonvanishing even-parity fields are constructed from master functions $h^\text{even}_{lm}$ and $a^\text{even}_{lm}$ in such a way that all 10 even-parity field equations will be satisfied after solving the master equations. Although, satisfaction of the fundamental field equations will further rely on the specific properties of the even-parity master equations. It can be shown that, when the master functions are defined according to Eqs.~\eqref{eq:heven}-\eqref{eq:ar}, the appropriate even-parity master equations have the same form as the odd-parity master equations,
\begin{align}
\label{eq:masterEven}
\left( \frac{d^2}{dr_*^2} + \omega^2 + \left[\begin{array}{cc} \alpha_l^\text{even} & \beta_l^\text{even} \\ \gamma_l^\text{even} & \sigma_l^\text{even} \end{array} \right] \right) \left[\begin{array}{c} h_{lm}^\text{even} \\ a_{lm}^\text{even} \end{array} \right] &= \left[\begin{array}{c} S_{lm}^\text{even} \\ Z_{lm}^\text{even} \end{array} \right] ,
\end{align}
but with alternate coefficients and source terms derived from the even-parity field equations, 
\begin{align}
&\alpha^\text{even}_l(r) = 2 f \left[ r^4 \left(r (3 M+\lambda  r)-2 Q^2\right)^2 \right]^{-1} \Big(Q^2 r^2 \left(21 M^2+16 \lambda  M r+2 (\lambda -1) \lambda r^2\right) \notag
\\&\qquad\qquad\qquad\qquad\qquad\qquad -r^3 \left(9 M^3+9 \lambda  M^2 r+3 \lambda ^2 M r^2+\lambda ^2 (\lambda +1) r^3\right)-2 Q^4 r (8 M+3 \lambda  r)+4 Q^6 \Big),
\\
&\beta^\text{even}_l(r) = \frac{8 f Q \left(-3 M^2 r+M \left(Q^2+3 r^2\right)+\lambda  (\lambda +2) r^3\right)}{r^2 \left(r (3 M+\lambda  r)-2 Q^2\right)^2} ,
\\
&\sigma^\text{even}_l(r) = -\frac{2 f \left(-2 Q^2 r^2 \left(9 M^2+8 \lambda  M r+(\lambda -1) \lambda  r^2\right)+2 Q^4 r (8 M+3 \lambda  r)+(\lambda +1) r^4 (3 M+\lambda  r)^2-4 Q^6\right)}{r^4 \left(r (3 M+\lambda  r)-2 Q^2\right)^2} ,
\\
&\gamma^\text{even}_l(r) = \frac{ \lambda f Q \left(-3 M^2 r+M \left(Q^2+3 r^2\right)+\lambda  (\lambda +2)
   r^3\right)}{r^2 \left(r (3 M+\lambda  r)-2 Q^2\right)^2} ,
\\
&S^\text{even}_{lm}(r) = \left[ r (3 M+\lambda  r)-2 Q^2 \right]^{-1} \bigg( r^2f\mathcal{Q}^{r}_{lm} + r^3 \mathcal{Q}^{rr}_{lm} - \frac{f(r(3M+\lambda r)-2Q^2)}{r} \mathcal{Q}^\sharp_{lm} - \frac{i\omega r^4 f}{\lambda+1} \mathcal{Q}^{tr}_{lm} + \frac{r^4f^3}{\lambda+1}\frac{d\mathcal{Q}^{tt}_{lm}}{dr} \notag
\\&\qquad\qquad\qquad -\frac{f^2 r \left(r^2 \left(12 M^2+3 (\lambda -3) M r+(\lambda -1) \lambda  r^2\right)+2 Q^2 r (5 r-8 M)+4 Q^4\right)}{(\lambda +1) \left(r (3 M+\lambda  r)-2 Q^2\right)} \mathcal{Q}^{tt}_{lm} +\frac{2Qr^2 f^2}{\lambda+1}\mathcal{J}^t_{lm} \bigg) ,
\\
&Z^\text{even}_{lm}(r) = \frac{r^2 Q}{2( r (3 M+\lambda  r)-2 Q^2)} \bigg( \frac{i \omega r f }{\lambda +1} \mathcal{Q}^{tr}_{lm} +\frac{f^2 \left(r \left(2 (\lambda +3) Q^2+\lambda  (\lambda +1) r^2\right)-M \left(2 Q^2+(\lambda +3) r^2\right)\right)}{r(\lambda +1) \left(r (3 M+\lambda  r)-2 Q^2\right)} \mathcal{Q}^{tt}_{lm} - \mathcal{Q}^{rr}_{lm} \notag
\\&\;\;\;\;  -\frac{f}{r} \mathcal{Q}^{r}_{lm} - \frac{f^3 r }{\lambda +1} \frac{d \mathcal{Q}^{tt}_{lm}}{dr}  \bigg) + \frac{r^2f^2}{2(\lambda+1)}\frac{d\mathcal{J}^t_{lm}}{dr} - \frac{i\omega r^2}{2(\lambda+1)}\mathcal{J}^r_{lm} -\frac{f\left(Q^4+Q^2 r(3r-4M)+r^2 (M-r)(3M+\lambda r) \right)}{r (\lambda+1) \left( r(3M+\lambda r)-2Q^2 \right)}\mathcal{J}^t_{lm} .
\end{align}
Additional master function properties, such as inverse master function relationships, are described in Appendix~\ref{sec:AppB}.

\section{Numerical calculations}
\label{sec:num}

\subsection{Circular orbital motion}
\label{sec:orb}

Here we demonstrate the techniques presented above by developing a computational model to describe gravitational waves emitted during the quasicircular inspiral of a charged compact object into a more massive charged black hole. This computational model is based on numerical calculations to determine the nonhomogeneous solution of the master equations with source terms representing a charged point mass. Therefore, to generate appropriate source terms we require a precise description of the orbital motion.

Accelerated motion of a point charge under the influence of the Lorentz force in RN spacetime can be described via an effective potential, $V_\text{eff}$~\cite{PuglQuevRuff11},
\begin{align}
&V_\text{eff}(r) = \frac{qQ}{r}+\mu\sqrt{f \left( 1+\frac{\mathcal{L}^2}{r^2} \right)} .
\end{align}
For circular orbits, the specific angular momentum, $\mathcal{L}$, is determined by finding a local minimum of $V_\text{eff}$ (evaluated at the orbital radius $r=r_p$) and solving for $\mathcal{L}$ in terms of $r_p$,
\begin{align}
& 0 = \frac{\partial V_\text{eff}}{\partial r} \, \bigg|_{r=r_p},
\\
& \mu \, \mathcal{L} = \frac{r_p \sqrt{ q^2 Q^2 r_p^2 f_p -2\mu^2 (Q^2-M r_p)(2 Q^2-3 M r_p+r_p^2) -q Q r_p^2 f_p\sqrt{q^2 Q^2+4 \mu^2(2 Q^2-3 M r_p+r_p^2)}}}{\sqrt{2}(2 Q^2-3 M r_p+r_p^2)} ,
\end{align}
where functions of $r$ with a $p$ subscript (such as $f_p$) are evaluated at $r = r_p$. The specific energy, $\mathcal{E}$, is $V_\text{eff}/\mu$ with the above value of $\mathcal{L}$ (and evaluated at $r=r_p$),
\begin{align}
&\mu \, \mathcal{E} = V_\text{eff}\big|_{r=r_p} .
\end{align}
It is useful to express the four-velocity, $u^\alpha$, in terms of the orbital energy and angular momentum \cite{Chan92},
\begin{align}
&u^t = \frac{1}{\mu f_p}\left( \mu \, \mathcal{E} -\frac{qQ}{r_p} \right),
\\
&u^\varphi = \frac{\mathcal{L}}{r_p^2}, 
\\
&u^r=u^\theta = 0.
\end{align}
The angular speed, $\Omega$, is the derivative of $\varphi_p$ with respect to $t$,
\begin{align}
\label{eq:Omega}
\Omega &= \frac{d\varphi_p}{dt} = \frac{u^\varphi}{u^t} .
\end{align}
The angular speed also serves as the fundamental angular frequency whose harmonics appear in the field equations,
\begin{align}
\label{eq:freqs}
& \omega_m = m \, \Omega ,
\end{align}
where $m$ is the spherical harmonic index that is introduced into the source modes during separation of variables.
The innermost stable circular orbit (ISCO) occurs at a particular orbital radius, $r_\text{ISCO}$, where the effective potential exhibits an inflection point,
\begin{align}
& 0 = \frac{\partial^2 V_\text{eff}}{\partial r^2} \bigg|_{r=r_p=r_\text{ISCO}} .
\end{align}
When $r_p$ slowly decreases to a value smaller than $r_\text{ISCO}$ during the inspiral, the smaller binary component will rapidly plunge into the black hole. See Appendix~\ref{sec:sources} for details about how these orbital properties are used to calculate the source terms driving the master equations.

\subsection{Solving the master equations}

The amplitudes of the nonhomogeneous master functions inform our model regarding the dissipation rate of the inspiral and its associated gravitational wave signature (see Sec.~\ref{sec:results}). We find those amplitudes by solving the nonhomogeneous master equations (Eqs.~\eqref{eq:masterOdd} and \eqref{eq:masterEven}) for each mode. Because the odd- and even-parity master equations have the same form, we omit the ``odd'' and ``even'' superscripts in Eqs.~\eqref{eq:masterOdd} and \eqref{eq:masterEven},
\begin{align}
\label{eq:master}
\left( \frac{d^2}{dr_*^2} + \omega_m^2 + \left[\begin{array}{cc} \alpha_l & \beta_l \\ \gamma_l & \sigma_l \end{array} \right] \right) \left[\begin{array}{c} h_{lm} \\ a_{lm} \end{array} \right] &= \left[\begin{array}{c} S_{lm} \\ Z_{lm} \end{array} \right] .
\end{align}
One property shared by the homogeneous ODE coefficients is that they all vanish approaching $r_*=\pm \infty$,
\begin{align}
\underset{r_* \rightarrow \pm \infty\;\;\;\;\;\;}{\text{lim}\;\, \alpha_l} = \underset{r_* \rightarrow \pm \infty\;\;\;\;\;\;}{\text{lim}\;\, \beta_l} =\underset{r_* \rightarrow \pm \infty\;\;\;\;\;\;}{\text{lim}\;\, \gamma_l} =\underset{r_* \rightarrow \pm \infty\;\;\;\;\;\;}{\text{lim}\;\, \sigma_l} = \;\; 0.
\end{align}
This property, in the context of Eq.~\eqref{eq:master}, requires that $h_{lm}$ and $a_{lm}$ behave as traveling waves near the event horizon ($r-r_+\ll M$) and in the far-field ($r \gg |\omega_m|^{-1}$). 

For point particles, the radial source terms involve Dirac delta functions
\begin{align}
\label{eq:source}
\left[\begin{array}{c} S_{lm} \\ Z_{lm} \end{array} \right] = \left[\begin{array}{c} B_{lm} \\ D_{lm} \end{array} \right]\delta(r-r_p) + \left[\begin{array}{c} F_{lm} \\ H_{lm} \end{array} \right] \delta'(r-r_p),
\end{align}
where $B_{lm}$, $D_{lm}$, $F_{lm}$, and $H_{lm}$ are constants determined by the orbital characteristics (see Appendix~\ref{sec:sources}), and a prime denotes differentiation with respect to $r$.

Equation \eqref{eq:master} has four independent homogeneous solutions. We denote each independent homogeneous solution with a superscript that implies certain boundary behavior. The two ``outgoing'' homogeneous solutions propagate toward $r_*=+\infty$ when $r\gg |\omega_m|^{-1}$,
\begin{align}
\label{eq:out0}
&\left[ \begin{array}{c} h_{lm}^{0+} \\ a_{lm}^{0+} \end{array} \right]_{r_* \rightarrow +\infty} \simeq e^{+i\omega_m r_*} \left[ \begin{array}{c} 1 \\ 0 \end{array} \right] , 
\\
\label{eq:out1}
& \left[ \begin{array}{c} h_{lm}^{1+} \\ a_{lm}^{1+} \end{array} \right]_{r_* \rightarrow +\infty} \simeq e^{+i\omega_m r_*} \left[ \begin{array}{c} 0 \\ 1 \end{array} \right] .
\end{align}
The two ``downgoing'' homogeneous solutions propagate toward $r_*=-\infty$ when $r-r_+ \ll M$,
\begin{align}
\label{eq:down0}
&\left[ \begin{array}{c} h_{lm}^{0-} \\ a_{lm}^{0-} \end{array} \right]_{r_* \rightarrow -\infty} \simeq e^{-i\omega_m r_*} \left[ \begin{array}{c} 1 \\ 0 \end{array} \right] , 
\\
\label{eq:down1}
&\left[ \begin{array}{c} h_{lm}^{1-} \\ a_{lm}^{1-} \end{array} \right]_{r_* \rightarrow -\infty} \simeq e^{-i\omega_m r_*} \left[ \begin{array}{c} 0 \\ 1 \end{array} \right] .
\end{align}
While other sets of homogeneous solutions are available, our chosen basis (Eqs.~\eqref{eq:out0}-\eqref{eq:down1}) is convenient for finding the retarded solution (our basis appropriately preserves causality). We expand each homogeneous solution in a power series near the boundaries to generate initial values for numerical integration of Eq. \eqref{eq:master}, see \cite{ZhuOsbu18}. These numerical integrations determine the global homogeneous solutions. 

Because the source terms involve Dirac delta functions (with no more than one derivative), the nonhomogeneous solution can be expressed as a piecewise function of homogeneous solutions,
\begin{align}
\label{eq:inhomo}
&\left[\begin{array}{c} h_{lm} \\ a_{lm} \end{array} \right] = \left( C_{lm}^{0+} \left[ \begin{array}{c} h_{lm}^{0+} \\ a_{lm}^{0+} \end{array} \right] + C_{lm}^{1+} \left[ \begin{array}{c} h_{lm}^{1+} \\ a_{lm}^{1+} \end{array} \right]  \right) \Theta(r-r_p)  +\left( C_{lm}^{0-} \left[ \begin{array}{c} h_{lm}^{0-} \\ a_{lm}^{0-} \end{array} \right] + C_{lm}^{1-} \left[ \begin{array}{c} h_{lm}^{1-} \\ a_{lm}^{1-} \end{array} \right]  \right) \Theta(r_p-r) ,
\end{align}
where $\Theta$ is the Heaviside step function. The normalization coefficients, $C_{lm}^{j\pm}$, are determined by enforcing a ``jump condition'' motivated by substituting Eq.~\eqref{eq:inhomo} into Eq.~\eqref{eq:master}. Through that procedure, a linear system (involving the Wronskian matrix) determines the normalization coefficients,
\begin{align}
&\left[\begin{array}{cccc} h_{lm}^{0+} & h_{lm}^{1+} & h_{lm}^{0-} & h_{lm}^{1-} 
\\ a_{lm}^{0+} & a_{lm}^{1+} & a_{lm}^{0-} & a_{lm}^{1-} 
\\ \partial_{r_*} h_{lm}^{0+} & \partial_{r_*} h_{lm}^{1+} & \partial_{r_*} h_{lm}^{0-} & \partial_{r_*} h_{lm}^{1-}
\\ \partial_{r_*} a_{lm}^{0+} & \partial_{r_*} a_{lm}^{1+} & \partial_{r_*} a_{lm}^{0-} & \partial_{r_*} a_{lm}^{1-}  \end{array} \right]_{r=r_p} \; \left[\begin{array}{c} C_{lm}^{0+} \\ C_{lm}^{1+} \\ -C_{lm}^{0-} \\ -C_{lm}^{1-} \end{array} \right]  = \frac{1}{r_p^3 f_p^2} \left[\begin{array}{c} r_p^3 F_{lm} \\ r_p^3 H_{lm} \\ r_p^3 f_p B_{lm}-2(Q^2-Mr_p) F_{lm} \\ r_p^3 f_p D_{lm}-2(Q^2-Mr_p) H_{lm} \end{array} \right] ,
\label{eq:wron}
\end{align}
where all radial functions have been evaluated at $r=r_p$.

Given $Q$, $q$, and $r_p$, our numerical approach to the above section is summarized in the following list:
\begin{enumerate}[(1)]
\item
A spherical harmonic ($l$, $m$) mode is chosen.
\begin{enumerate}[a.]
\item
$l$ is restricted to the range $l\ge 1$.
\item
$m$ is restricted to the range $1 \le m \le l$.
\item
If $l+m$ is even, we use the even-parity equations.
\item
If $l+m$ is odd, we use the odd-parity equations.
\end{enumerate}
\item
A custom \textsc{Python} function is used to generate initial values for numerical integration of the homogeneous solutions through series expansions (see \cite{ZhuOsbu18}).
\item
The homogeneous version of Eq. \eqref{eq:master} is integrated numerically using \emph{scipy.integrate.odeint} in \textsc{Python} (with accuracy tolerance = $10^{-13}$) for each set of initial values.
\begin{enumerate}[a.]
\item
The initial position $r_i = 30/|\omega_m| + 10M$ is used for the solutions described by Eqs.~\eqref{eq:out0} and \eqref{eq:out1}.
\item
The initial position $r_i = r_+ + 10^{-3} M$ is used for the solutions described by Eqs.~\eqref{eq:down0} and \eqref{eq:down1}.
\item
The final position $r_f = r_p$ is used for all homogeneous integrations.
\end{enumerate}
\item
The nonhomogeneous solution is found using Eq. \eqref{eq:inhomo}.
\begin{enumerate}[a.]
\item
The Wronskian matrix is generated using the homogeneous solutions evaluated at $r=r_p$.
\item
The source vector is generated using the orbital characteristics implied by $Q$, $q$, and $r_p$ (see Appendix \ref{sec:sources}).
\item
The normalization coefficients are calculated by solving Eq. \eqref{eq:wron} using \emph{numpy.linalg.solve} in \textsc{Python}.
\end{enumerate}
\item
Steps (1)-(4) are repeated for all $l$ and $m$ values through $l_\text{max}=25$ (which we have determined to be a sufficient level of convergence).
\end{enumerate}

\subsection{Radiation reaction}

The average rate of radiative energy dissipation, $\langle \dot{E} \rangle$, is quantified through analysis of gravitational and electromagnetic wave propagation towards the far field and into the event horizon, see~\cite{ZhuOsbu18}. The normalization coefficients for the nonhomogeneous solutions of the master equations quantify the combined gravitational and electromagnetic energy flux, 
\begin{align}
\langle \dot{E} \rangle = \sum_{l=1}^{l_\text{max}} \sum_{m=1}^l \frac{\omega_m^2}{32\pi}\bigg( (l+2)(l+1)l(l-1)\Big(\big|C^{0+}_{lm}\big|^2+\big|C^{0-}_{lm}\big|^2\Big) + 16\, l(l+1) \Big(\big|C^{1+}_{lm}\big|^2+\big|C^{1-}_{lm}\big|^2\Big) \bigg) .
\end{align}
The dissipative component of the self-force~\cite{MinoSasaTana97,QuinWald97,Pois11} accounts for this radiative energy loss by causing the orbital energy to decrease at the same rate. To avoid technical challenges associated with calculating the self-force directly, we incorporate radiation reaction through an adiabatic approximation based on energy balance arguments,
\begin{align}
\label{eq:fluxBalance}
\mu \frac{d\mathcal{E}}{dt} &= - \langle \dot{E} \rangle .
\end{align}
By allowing the orbital radius, $r_p$, to slowly decrease (the quasicircular approximation), we can use Eq.~\eqref{eq:fluxBalance} to generate equations of motion by calculating $\langle \dot{E} \rangle$ from the nonhomogeneous master equations with different circular orbits over a spectrum of orbital radii,
\begin{align}
\label{eq:EOMr}
\frac{d r_p}{dt} &= -\dfrac{\langle \dot{E} \rangle}{\mu} \left( \dfrac{\partial \mathcal{E}}{\partial r_p} \right) ^{-1} ,
\\
\label{eq:EOMphi}
\frac{d\varphi_p}{dt} &= \Omega .
\end{align}
Recall that $\mathcal{E}$ and $\Omega$ are known functions of $r_p$ according to Sec.~\ref{sec:orb}. Therefore, the instantaneous velocities described by Eqs.~\eqref{eq:EOMr} and~\eqref{eq:EOMphi} depend only on $r_p$. It is then straightforward to choose an initial orbital radius and integrate Eqs. \eqref{eq:EOMr} and \eqref{eq:EOMphi} numerically to find $r_p$ and $\varphi_p$ as functions of $t$.

One interesting challenge is that, while $\mathcal{E}$ and $\Omega$ are known exactly in terms of $r_p$, $\langle \dot{E} \rangle$ must be calculated numerically from the master equations at each new orbital radius. It would be inefficient to directly couple the equations of motion to the master equations because that would involve re-solving Eq.~\eqref{eq:master} for every $l$ and $m$ at each function evaluation during numerical integration of Eqs.~\eqref{eq:EOMr} and~\eqref{eq:EOMphi}. Instead, we precompute $\langle \dot{E} \rangle$ for a dense set of orbital radii and perform an interpolation. Then this interpolant can be evaluated rapidly during the inspiral evolution. We use the same interpolation strategy as~\cite{ZhuOsbu18}, except with twice the density of $r_p$ samples. This interpolation must be performed separately for each pair of $Q/M$ and $q/\mu$ values under investigation.

\end{widetext}

\section{Results}
\label{sec:results}

\subsection{Inspirals}

After numerically generating dissipation data throughout a range of orbital radii for numerous charge-to-mass-ratio pairs ($Q/M$ and $q/\mu$), we solved Eqs.~\eqref{eq:EOMr}-\eqref{eq:EOMphi} to determine inspiral trajectories. Figure~\ref{fig:inspirals} depicts various trajectories with $\epsilon = 0.1$ (which stretches the limitations of our small mass-ratio approximation, but provides an effective illustration). When comparing different charge-to-mass ratios, we intentionally choose different initial orbital radii that minimize relative orbital dephasing (which also minimizes waveform dephasing). Minimization of dephasing is accomplished by matching the initial orbital frequencies of the binary systems under comparison. It is straightforward to fix one initial orbital radius and use root finding (for the initial frequency difference according to Eq.~\eqref{eq:Omega}) to determine the other initial orbital radius. As the contrasting binaries evolve, their disparate properties cause a relative drift of their orbital frequencies (and subsequent dephasing) over time.

The uppermost frame of Fig.~\ref{fig:inspirals} is representative of previously existing (noncharged) adiabatic quasicircular inspiral models for comparison. The other cases investigate how charging one or both bodies affects inspiral dynamics. It is apparent from Fig.~\ref{fig:inspirals} that the inspiral decays more rapidly when electric charge is present on either the smaller or larger binary component. One noteworthy result is that the system decays more rapidly when only the smaller body is charged compared to when only the central black hole is charged (at the same charge-to-mass-ratios). When the two bodies possess like charge-to-mass-ratios, the system decays more slowly than the neutral case. Finally, when the two bodies have opposite charge-to-mass-ratios the system decays more rapidly than any other case. These outcomes for like and opposite charges harmonize with certain nonrelativistic intuition (attraction would increase centripetal acceleration and, according to the Larmor formula, intensify dissipation, while repulsion would have the opposite effect). Building on these observations, we seek to determine the necessary conditions (minimum amount of charge) for which charged binary components would impact gravitational wave astronomy.

\subsection{Dephasing}

\begin{figure}
\includegraphics[width=2.98in]{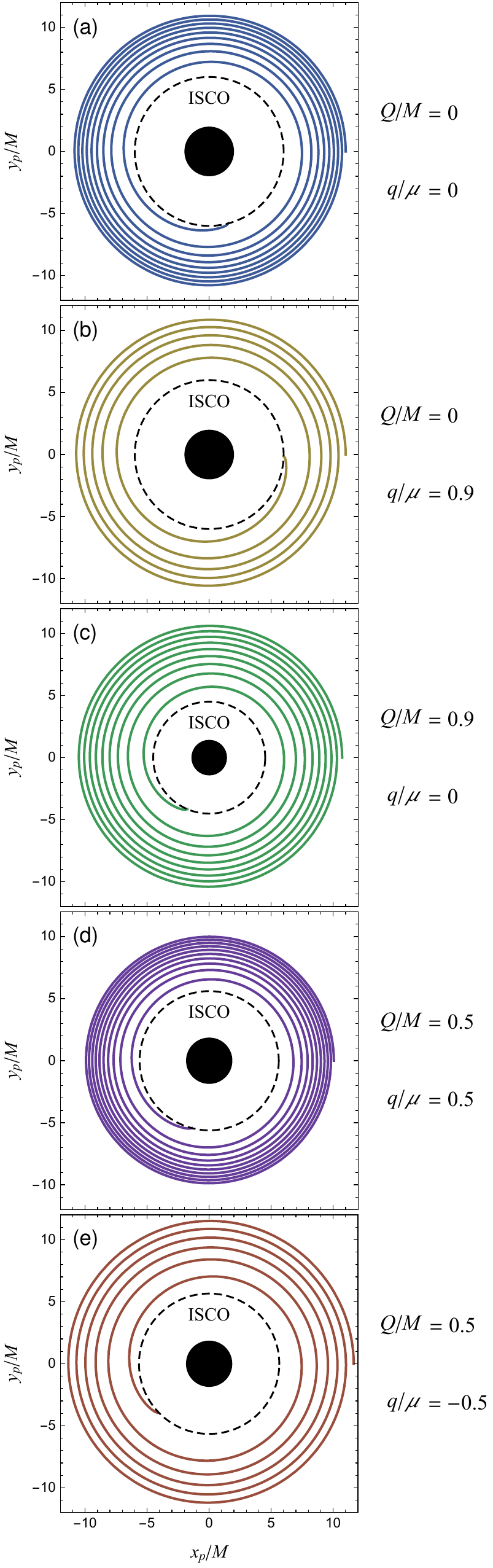}
\end{figure}

\begin{figure}
\includegraphics[width=3.4in]{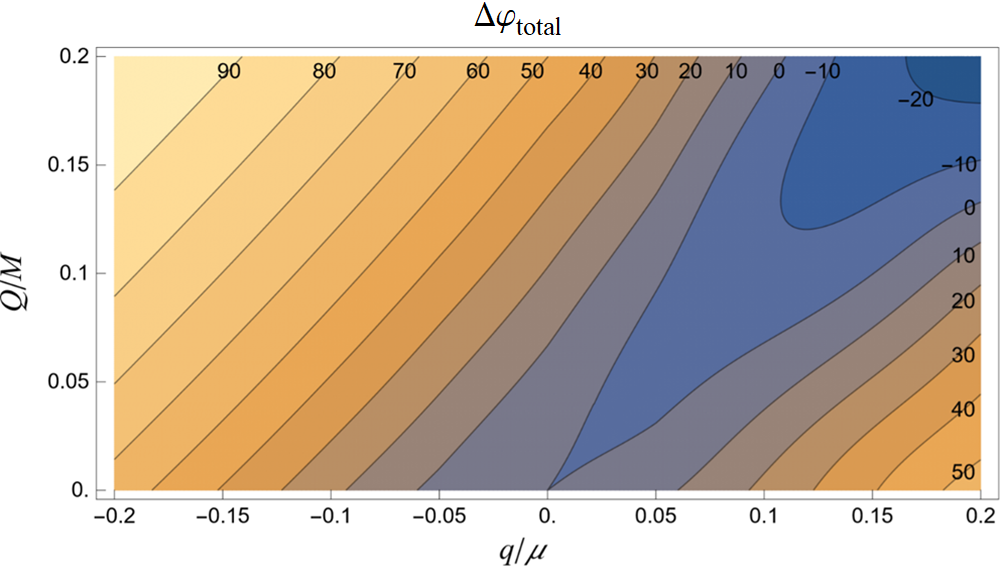}
\caption{\label{fig:contours} This contour plot depicts lines of constant $\Delta\varphi_\text{total}$ for numerous combinations of $q/\mu$ (horizontal axis) and $Q/M$ (vertical axis). Each position on the $q/\mu$--$Q/M$ plane represents a charged inspiral that was compared with a neutral inspiral to determine their accumulated phase difference ($\Delta\varphi_\text{total}$). For this plot we have fixed the mass-ratio at a constant value of $\epsilon=0.01$ (changing $\epsilon$ shifts the values of each contour by a uniform factor, but has a negligible effect on the qualitative pattern). Notice that the most dramatic dephasing occurs when $-q/\mu=Q/M$ (the gradient is steepest in that direction from the origin). Conversely, the amount of dephasing has a negative correlation with charge-to-mass-ratio when $q/\mu=Q/M$, which is to say that binaries of like charge traverse fewer total revolutions than neutral binaries (with comparable initial conditions and the same overall duration). Figure~\ref{fig:smallCharge} further investigates these two cases, along with the $Q=0$ and $q=0$ cases, in the realm of smaller charge. }
\end{figure}

To quantify observational impact, we examine dephasing of charged binaries relative to noncharged binaries (the prevalent case). Initially, we quantify dephasing via the accumulated orbital phase difference: $\Delta\varphi_\text{total} \equiv \varphi_{q\ne 0}^{Q\ne 0}(t_\text{max}) - \varphi_{q=0}^{Q=0}(t_\text{max})$, where $t_\text{max}$ is a time shortly before plunge. Later we assess the waveforms directly (see Sec.~\ref{sec:waveforms}). To investigate a broad range of cases, we performed inspiral integrations for 45 pairs of charge-to-mass-ratios in the overlapping ranges $0\le Q/M \le 0.2$ and $-0.2 \le q/\mu \le 0.2$ (later we introduce an additional data set focusing on smaller charges). Throughout this subsection, we use the initial value $r_p(0)\simeq 20M$ as a rough guideline for appropriate conditions. Figure~\ref{fig:contours} depicts these data in the form of a contour plot featuring lines of constant $\Delta\varphi_\text{total}$. Many of our preliminary observations from Fig.~\ref{fig:inspirals} are affirmed with greater detail by Fig.~\ref{fig:contours}. As one example, notice that the steepest gradient of $\Delta\varphi_\text{total}$ in Fig.~\ref{fig:contours} is in the direction of opposite charge-to-mass-ratios ($-q/\mu =Q/M$), which is consistent with the rapid decay rate for opposite charges in Fig.~\ref{fig:inspirals}. Similarly, the like charge-to-mass-ratio direction in Fig.~\ref{fig:contours} involves a negative $\Delta\varphi_\text{total}$, which matches the slower decay rate for like charges in Fig.~\ref{fig:inspirals}.

\begin{figure}[b]
\caption{\label{fig:inspirals} Left: inspiral trajectories of the smaller binary component until ISCO. A larger mass-ratio, $\epsilon=0.1$ (near the upper limits of our approximation scheme), provides an accentuated illustration of how different charges affect the inspiral. The different starting orbital radii are chosen to minimize relative dephasing (which is appropriate for comparisons) by matching the initial orbital frequencies. Frame (a) shows the trajectory with two neutral bodies. Frame (b) shows the trajectory with charge-to-mass-ratios $Q/M=0$ and $q/\mu=0.9$. Radiation for this case has been studied  previously~\cite{DaviRuffTiom72}, but here we present the first inspiral calculations with a smaller body that is charged. It is perhaps unsurprising that frame (b) decays more rapidly than frame (a) because of the additional dissipation mechanism (electromagnetic radiation). A more interesting result follows from comparison of frames (b) and (c) (the latter, with $Q/M=0.9$ and $q/\mu=0$, has been studied previously~\cite{ZhuOsbu18}). We find that, while (b) and (c) both decay more rapidly than (a) (the neutral case), (b) decays more rapidly than (c). This suggests that, when only one component is charged, the charge-to-mass-ratio of the smaller body has a more dramatic effect than that of the larger body. Frames (d) and (e) represent the first inspiral calculations where both bodies are charged. Frame (d) shows the case of like charge-to-mass-ratios ($Q/M=0.5$ and $q/\mu=0.5$), which decays less rapidly than (a) (the neutral case). Finally, frame (e) demonstrates that the case of opposite charges ($Q/M=0.5$ and $q/\mu=-0.5$) involves that fastest decay of all (much faster comparatively if the sizes of all nonvanishing charge-to-mass-ratios were fixed across cases).}
\end{figure}

\begin{figure}
\includegraphics[width=3.4in]{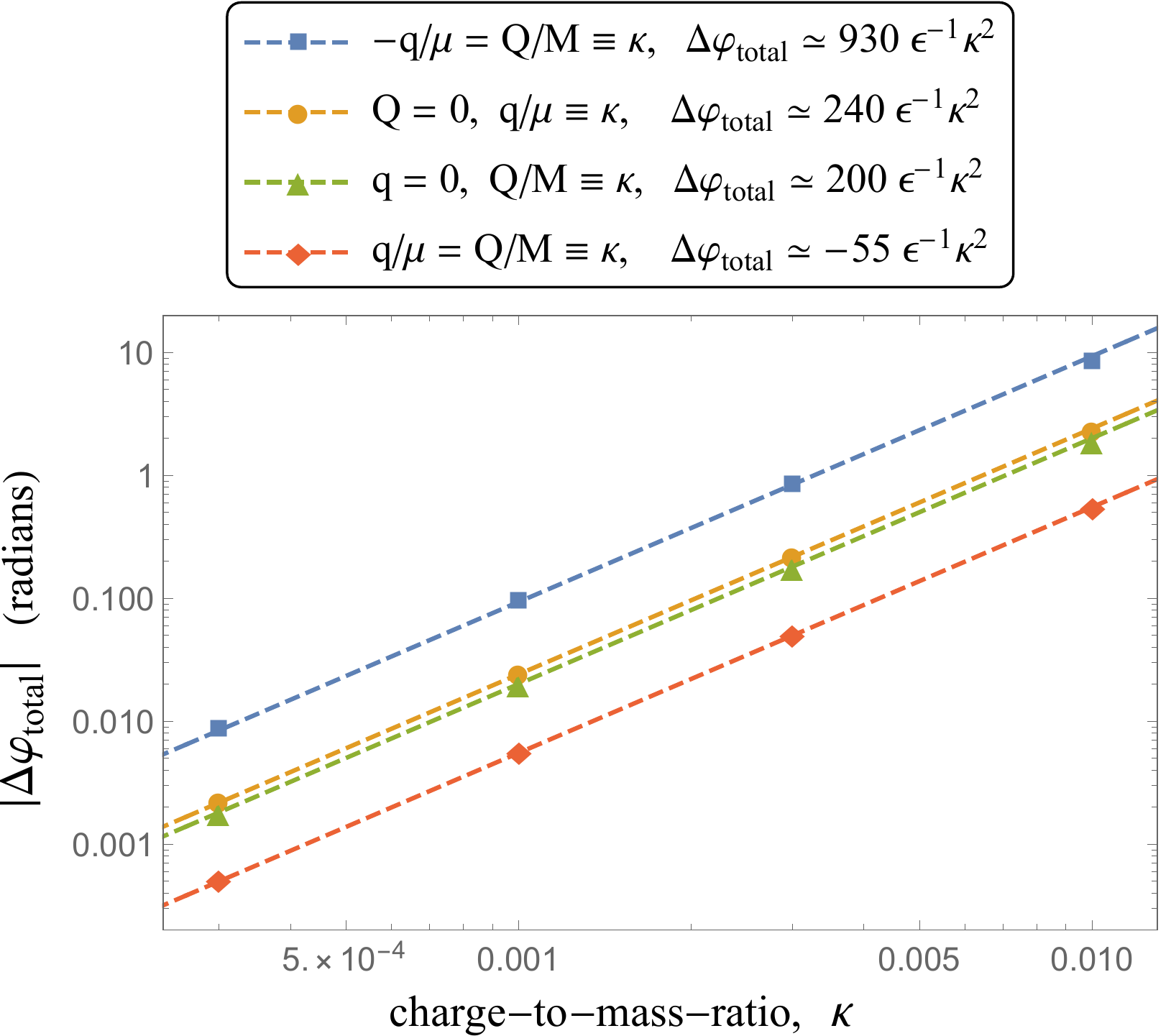}
\caption{\label{fig:smallCharge} To highlight cases with small amounts of charge, we plot $|\Delta \varphi_\text{total}|$ vs. charge-to-mass-ratio ($\kappa$) on a log-log scale for four different categories reflecting possible small charge scenarios (see legend for unique shapes/colors). Each data point represents a numerical calculation for the amount of dephasing (vertical axis) relative to a noncharged binary. In this plot the mass-ratio is fixed at $\epsilon=0.01$, but we performed similar calculations for a variety of mass-ratios. With the aid of least-squares fitting, we determined best-fit coefficients for power-law models (valid with small amounts of charge for each category) describing $\Delta \varphi_\text{total}$ as a function of charge-to-mass-ratio, $\kappa$, and mass-ratio, $\epsilon$. Approximate expressions for these fitted models are available in the legend, and their behaviors are depicted by the dashed lines. For the $q=0$ ($Q/M \ne 0$) category, our best-fit coefficients agree with past work~\cite{ZhuOsbu18}. The best-fit coefficients for the other three categories are new results. Although the definition of $\kappa$ is different for each category, in all four categories $\Delta \varphi_\text{total}$ depends quadratically on the appropriate charge-to-mass-ratio. Another (perhaps expected) result is the inverse dependence of $\Delta\varphi_\text{total}$ on the mass-ratio, $\epsilon$. Many of the qualitative results from Fig.~\ref{fig:contours}, such as the intensified dephasing for $-q/\mu =Q/M$, are confirmed quantitatively by this small charge analysis. }
\end{figure}

Because the very existence of charged compact bodies is speculative, next we highlight cases involving small charge-to-mass-ratios (as those may be the most likely to exist). For this small charge analysis we consider four categories: $Q=0$ (with $q/\mu\ne 0$), $q=0$ (with $Q/M\ne 0$), $q/\mu=Q/M$, and $-q/\mu=Q/M$. For each category we numerically calculated $\Delta\varphi_\text{total}$ for a variety of small charge-to-mass-ratios. Figure~\ref{fig:smallCharge} demonstrates that, in all four categories, $\Delta\varphi_\text{total}$ depends quadratically on the charge-to-mass-ratio (but with unique coefficients for each category) in the realm of small charge. Another (perhaps expected) result is the inverse dependence of $\Delta\varphi_\text{total}$ on the mass-ratio, $\epsilon$. With the aid of least-squares fitting, we determined best-fit coefficients for $\Delta\varphi_\text{total}$ as a function of the charge-to-mass-ratio for each category (also shown in Fig.~\ref{fig:smallCharge}). These approximate functions for $\Delta\varphi_\text{total}$ in the small charge realm are in qualitative agreement with the broader (but less quantitative) picture of Fig.~\ref{fig:contours}. One of these shared predictions, the intensified dephasing in cases with $-q/\mu =Q/M$, motivates our choice of system for exploring gravitational wave signals directly.

\begin{figure}
\includegraphics[width=3.4in]{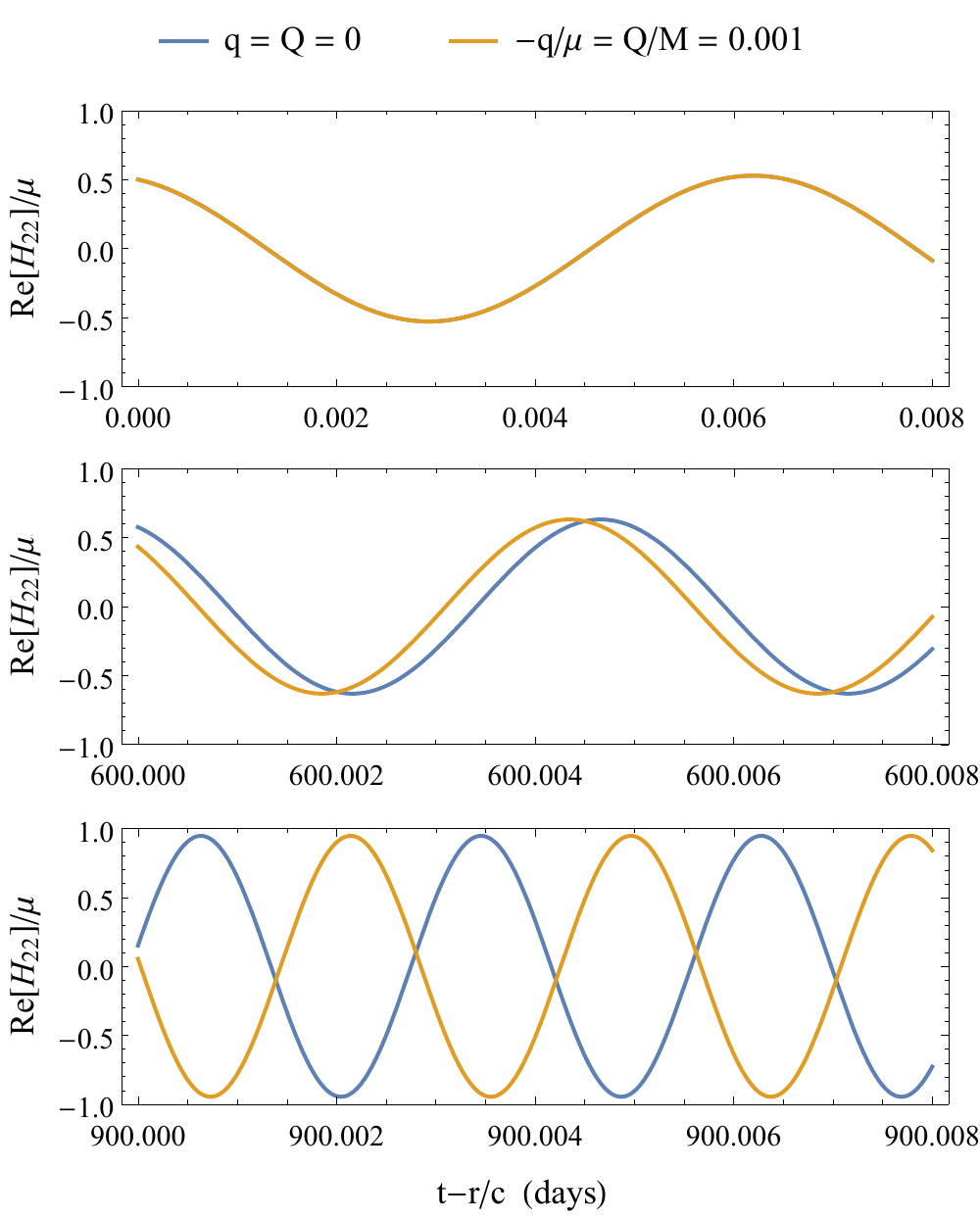}
\caption{\label{fig:waveforms} Quadrupole waveforms for two extreme mass-ratio inspirals with $M=10^6 M_\odot$ and $\mu=10 M_\odot$. The blue waveform involves noncharged components while the orange waveform involves bodies with opposite charge-to-mass-ratios: $-q/\mu=Q/M=0.001$. The noncharged inspiral has an initial orbital radius $r_p(0)=11 M$, while the charged inspiral has a slightly different initial orbital radius chosen to match the initial frequencies of the two waveforms (the same matching has been performed to ensure consistency in all prior comparisons). Notice that even this relatively small charge-to-mass-ratio (1/1000) is sufficient for gravitational wave signals involving charged bodies to entirely dephase from those of a noncharged binary system.}
\end{figure}

\subsection{Waveforms}
\label{sec:waveforms}

Predicting the nature of gravitational wave signals will require interfacing our inspiral model with an appropriate waveform generation scheme. Although simple qualitative approximations exist (such as kludge models), formally the waveform is encoded within the far-field (nonhomogeneous) metric perturbation. Fortunately, the same process for determining the energy dissipation rate (which forms the backbone of our inspiral scheme) also provides complex amplitudes describing the far-field metric perturbation~\cite{WarbOsbuEvan17},
\begin{align}
&h_+ - ih_\times = \frac{1}{r}\sum_{l=2}^{l_\text{max}}\sum_{m=-l}^l H_{lm}(t,r) \;_{-2}Y_{lm}(\theta,\varphi) ,
\\
&H_{lm}(t,r) \equiv \frac{C^{0+}_{lm}}{2}\sqrt{(l+2)(l+1)l(l-1)}\; e^{-im\varphi_p|_{t \rightarrow t-r}} ,
\end{align}
where $\,_{-2}Y_{lm}$ is a spin-weighted spherical harmonic. Similar to our treatment of the inspiral dissipation rate, we interpolate (using the same scheme and numerical data set as before) the waveform amplitudes across orbital radii, so that $C^{0+}_{lm}$ is a function of $r_p|_{t\rightarrow t-r}$ (the replacement indicates evaluation at the retarded time: $t\rightarrow t-r$). These waveforms are not exact because each snapshot assumes a fixed orbital radius (in reality the orbital radius is decreasing), however, this "evolving snapshots" technique provides an accurate approximation for the waveform~\cite{WarbOsbuEvan17}.

To investigate the differences between waveforms from charged and noncharged binaries, we consider two extreme mass-ratio inspirals (one with charged bodies, the other with noncharged bodies) with comparable properties. The two inspiraling systems involve the same pairs of masses for the larger and smaller binary components that we chose to represent typical LISA sources: $M=10^6 M_\odot$ and $\mu = 10 M_\odot$ (therefore $\epsilon=10^{-5}$). Following evidence from Figs.~\ref{fig:contours} and \ref{fig:smallCharge}, for the charged system we calculate waveforms in the opposite charge-to-mass-ratio case (which maximizes contrast with the noncharged system). Figure~\ref{fig:waveforms} demonstrates that an opposite charge-to-mass-ratio of $-q/\mu = Q/M = 0.001$ is sufficient for gravitational wave signals involving charged bodies to entirely dephase from those of a noncharged binary system. 

\section{Conclusions and Future Work}

In this work we presented an improved Reissner-Nordström perturbation formalism based on newly enhanced (compared to prior work~\cite{Zeri74,Monc74a,Monc74b,Monc75}) master equations and master functions. Our method involves perturbatively expanding the spacetime metric and electromagnetic potential four-vector in the vicinity of a Reissner-Nordström black hole. After applying separation of variables via vector and tensor spherical harmonic decompositions, we defined gravitational and electromagnetic master functions with convenient properties. Our master function framework allows for perturbations from arbitrary distributions of electrically charged matter and, unlike aforementioned nonhomogeneous master function formalisms, is well suited for time-domain calculations. 

We pursued an application of our formalism involving gravitational wave source modeling by considering intermediate and extreme mass-ratio inspirals of a charged compact object into a Reissner-Nordström black hole. Other than passing references to prior work~\cite{Wald74,Ruff77,Puns98,RujuETC90,CardETC16}, we did not investigate further how the two bodies may have become charged in the first place. Instead, we sought to quantify an amount of charge (on one or both bodies) for which existing noncharged theoretical models would be insufficient for gravitational wave data analysis after a hypothetical detection involving a charged two-body source. In pursuit of this goal, we applied our master function formalism to a variety of two-body systems with various charges and mass-ratios. For each pair of charges ($q$ and $Q$), we solved our master equations numerically across multipole modes with a point source (a charged point mass) in circular motion. We used the nonhomogeneous master function amplitudes to quantify the rate of radiative energy decay as a function of orbital radius (by re-solving the master equations with a new source for each radius). This dissipation rate facilitated approximate equations of motion governing the slowly decreasing orbital radius that is characteristic of quasicircular inspirals.

Through these calculations, we observed that the case of opposite charge-to-mass-ratios ($-q/\mu = Q/M$) involved the most significant dephasing compared to the neutral case. We also quantified lesser dephasing for like charges ($q/\mu = Q/M$) and solo charges ($q=0$ or $Q=0$). By comparing gravitational waveforms for charged vs. neutral binaries, we determined that charge-to-mass-ratios greater than $\sim 10^{-3}$ can involve total dephasing for sources accessible by LISA (we examined $\epsilon \simeq 10^{-5}$). Based on these results, we believe it is plausible that LISA observations could be used to quantify whether (and how much) compact binary components are charged (or perhaps an upper limit could be established).

Future applications of this framework include noncircular orbital motion and/or self-force calculations. Bound eccentric orbital motion is one important~\cite{Hopm05} noncircular case that would introduce an additional fundamental frequency and an associated spectrum of discrete harmonics, but otherwise should involve similar techniques. Other noncircular cases, such as hyperbolic encounters or many body systems, would benefit from the time-domain applicability of our framework. Calculating the gravitational and electromagnetic self-forces in Reissner-Nordström spacetime would require a careful regularization of the field near the small body. Such techniques have been applied for the scalar self-force~\cite{Bara00,Burk01,CastVegaWard18}, but not for the gravitational and electromagnetic self-forces in Reissner-Nordström spacetime (but preliminary work has been conducted~\cite{ZimmPois14,Zimm15,Linz14}). If the appropriate self-forces were calculated, they could be applied to investigate the mechanisms behind cosmic censorship~\cite{ZimmETC13,SorcWald17}.

Other potential directions might involve generalization of the background environment to include additional effects. One obvious enhancement would be to consider a central black hole that is both charged and rotating (perturbations of Kerr-Newman spacetime). This is appropriate because some models require spinning black holes to explain how charge might accumulate~\cite{Wald74,Ruff77,Puns98}. However, past efforts have been unsuccessful in applying separation of variables to Kerr-Newman perturbations~\cite{Chan92b} (and so we believe this direction to be especially challenging). There are also other effects that could impact the environment surrounding the black hole. One example involves considering that a nearby plasma could radiate upon acceleration of its constituent particles during the inspiral and merger process~\cite{Zhan16}. Another example is the prediction that charged black holes could be surrounded by an oppositely charged magnetosphere~\cite{Puns98}, which would likely affect the dynamics of the smaller body and its gravitational wave signature.

\acknowledgments
J. B. and T. O. thank Ruomin Zhu for providing comparison data. J. B. thanks Erin Bonning for helpful discussion regarding astrophysical black hole mass distributions. J. B. and T. O. gratefully acknowledge support from the SURE-Oxford program. J. B. gratefully acknowledges support from the New York Space Grant, Academic Affairs budget of Oxford College, and the APS Future of Physics Days Travel Grant.

\begin{widetext}

\begin{appendix}

\section{Source terms}
\label{sec:sources}

The individual spherical harmonic modes that make up the source terms are calculated by leveraging the orthogonality of the angular basis functions \cite{MartPois05},
\begin{align}
\label{eq:sourceA}
&\mathcal{J}^{a}_{lm}\,e^{-i\omega t} = 4\pi \int J^{a} \bar{Y}^{lm} d\Omega ,
\\
&\mathcal{J}^{\sharp}_{lm}\,e^{-i\omega t} = \frac{4\pi r^2}{l(l+1)}\int J^{A} \bar{Y}^{lm}_{A} d\Omega ,
\\
&\mathcal{J}^{\flat}_{lm}\,e^{-i\omega t} = \frac{4\pi r^2}{l(l+1)}\int J^{A} \bar{X}^{lm}_{A} d\Omega ,
\\
&\mathcal{Q}^{ab}_{lm}\,e^{-i\omega t} = 8\pi \int T^{ab}_\text{matter} \bar{Y}^{lm} d\Omega ,
\\
&\mathcal{Q}^{a}_{lm}\,e^{-i\omega t} = \frac{16\pi r^2}{l(l+1)}\int T^{aB}_\text{matter} \bar{Y}^{lm}_B d\Omega ,
\\
&\mathcal{Q}^{\flat}_{lm}\,e^{-i\omega t} = 8\pi r^2\int T^{AB}_\text{matter}\Omega_{AB} \bar{Y}^{lm} d\Omega ,
\\
&\mathcal{Q}^{\sharp}_{lm}\,e^{-i\omega t} = \frac{32\pi r^4}{(l+2)(l+1)l(l-1)}\int T^{AB}_\text{matter} \bar{Y}^{lm}_{AB} d\Omega ,
\\
&\mathcal{P}^{a}_{lm}\,e^{-i\omega t} = \frac{16\pi r^2}{l(l+1)}\int T^{aB}_\text{matter} \bar{X}^{lm}_B d\Omega ,
\\
&\mathcal{P}_{lm}\,e^{-i\omega t} = \frac{16\pi r^4}{(l+2)(l+1)l(l-1)}\int T^{AB}_\text{matter} \bar{X}^{lm}_{AB} d\Omega .
\label{eq:sourceZ}
\end{align}
Similar to Sec.~\ref{sec:multipole}, the sources are represented as if they exhibit sinusoidal time-dependence, but Eqs.\eqref{eq:sourceA}-\eqref{eq:sourceZ} are easily generalizable for arbitrary time-dependence via Fourier transform. For a charged point mass, the sources are described by Eqs.~\eqref{eq:J} and \eqref{eq:Tmat}. It is straightforward to find expressions for the multipole modes of a point particle source in circular motion,
\begin{align}
&\mathcal{J}^r_{lm} = \mathcal{Q}^{tr}_{lm} = \mathcal{Q}^{rr}_{lm} = \mathcal{Q}^{r}_{lm} = \mathcal{P}^{r}_{lm} = 0, 
\\
&\mathcal{J}^{t}_{lm} = \frac{4\pi q}{r_p^2} \delta(r-r_p)\, Y^{lm}\left(\frac{\pi}{2},0\right),
\\
&\mathcal{J}^{\sharp}_{lm} = \frac{4\pi q u^{\varphi}}{l(l+1)u^t} \delta(r-r_p)\, Y^{lm}_\varphi \left(\frac{\pi}{2},0\right) ,
\\
&\mathcal{J}^{\flat}_{lm} = \frac{4\pi q u^\varphi}{l(l+1) u^t} \delta(r-r_p)\, X^{lm}_\varphi \left(\frac{\pi}{2},0\right) ,
\\
&\mathcal{Q}^{tt}_{lm} = \frac{8 \pi \mu u^t}{r_p^2} \delta(r-r_p)\, Y^{lm}\left(\frac{\pi}{2},0\right),
\\
&\mathcal{Q}^{t}_{lm} = \frac{16 \pi \mu u^\varphi}{l(l+1)} \delta(r-r_p)\, Y^{lm}_\varphi \left(\frac{\pi}{2},0\right) ,
\\
&\mathcal{Q}^\flat_{lm} = \frac{8 \pi \mu (u^\varphi)^2}{u^t} \delta(r-r_p)\, Y^{lm} \left(\frac{\pi}{2},0\right) ,
\\
&\mathcal{Q}^\sharp_{lm} = \frac{32 \pi \mu r_p^2 (u^\varphi)^2}{(l+2)(l+1)l(l-1)u^t} \delta(r-r_p)\, Y^{lm}_{\varphi\varphi} \left(\frac{\pi}{2},0\right),
\\
&\mathcal{P}^t_{lm} = \frac{16 \pi \mu u^\varphi}{l(l+1)} \delta(r-r_p)\, X^{lm}_\varphi \left(\frac{\pi}{2},0\right) ,
\\
&\mathcal{P}_{lm} = \frac{16 \pi \mu r_p^2 (u^\varphi)^2}{(l+2)(l+1)l(l-1)u^t} \delta(r-r_p)\, X^{lm}_{\varphi\varphi} \left(\frac{\pi}{2},0\right) ,
\end{align}
where $\varphi$ subscripts indicate differentiation with respect to $\varphi$ (following~\cite{MartPois05}). These are the building blocks for the master equation sources that appear in Eq.~\eqref{eq:source}, with sets of odd-parity coefficients:
\begin{align}
&B_{lm}^\text{odd} = \frac{32\pi \mu f_p(Q^2-r_p^2)u^\varphi}{r_p^2(l+2)(l+1)l(l-1)}\sqrt{(l+m+1)(l-m)}\, Y^{l,m+1}\left(\frac{\pi}{2},0\right)\, ,
\\
&F_{lm}^\text{odd} = \frac{32\pi \mu r_p f_p^2 u^\varphi}{(l+2)(l+1)l(l-1)}\sqrt{(l+m+1)(l-m)}\, Y^{l,m+1}\left(\frac{\pi}{2},0\right)\, , 
\\
&D_{lm}^\text{odd} = \frac{4\pi q f_p u^\varphi}{l(l+1) u^t} \sqrt{(l+m+1)(l-m)}\, Y^{l,m+1}\left(\frac{\pi}{2},0\right) \, , 
\\
&H_{lm}^\text{odd} = 0 .
\end{align}
and even-parity coefficients:
\begin{align}
& B_{lm}^\text{even} = 8 \pi  f_p \left[ \lambda  (\lambda +1) r_p^3 u^t \left(r_p \left(3 M+\lambda  r_p\right)-2 Q^2\right)^2 \right]^{-1} \Bigg( \lambda  q Q f_p r_p^3 u^t \left(r_p \left(3 M+\lambda r_p\right)-2 Q^2\right) \notag
\\&\;\;\; - \mu  \Big[\lambda  \Big(f_p r_p^2 (u^t)^2 \left(12 M^2 r_p^2+Q^2 r_p \left(2 r_p-21 M\right)+8 Q^4\right)+r_p^4 (u^{\varphi})^2 \left(2 Q^2-3 M r_p\right) \left[r_p \left(2 (m^2-1) r_p-3 M\right)+2 Q^2\right]\Big) \notag
\\&\;\;\qquad\qquad\qquad +\lambda ^2 r_p^2 \Big(r_p^4 (u^{\varphi})^2 \left[r_p \left(r_p(1-m^2)+6 M\right)-4 Q^2\right]+f_p r_p^2 (u^t)^2 \left(r_p \left(5 M+r_p\right) - 4 Q^2\right)\Big) 
\\&\qquad\qquad\qquad\qquad\qquad\qquad +\lambda ^3 r_p^4 \left(f_p r_p^2 (u^t)^2+r_p^4 (u^{\varphi})^2\right) -\left(m^2-1\right) r_p^4 (u^{\varphi})^2 \left(2 Q^2-3 M r_p\right)^2 \Big] \Bigg) \, Y^{lm}\left( \frac{\pi}{2},0 \right) \, , \notag
\\
& F_{lm}^\text{even} = \frac{8 \pi  \mu  f_p^3 r_p^2 u^t}{(\lambda +1) \left(r_p
   (3 M+\lambda  r_p)-2 Q^2\right)} \, Y^{lm}\left(\frac{\pi}{2},0\right)\, ,
\\
& D_{lm}^\text{even} = 4 \pi f_p \Big[ (\lambda +1) r_p^3 \left(r_p \left(3 M+\lambda  r_p\right)-2 Q^2\right)^2 \Big]^{-1} \Bigg(\mu  Q f_p r_p u^t \Big[r_p^2 (6 M^2+3 (\lambda +1) M r_p+\lambda  (\lambda +2) r_p^2) +6 Q^4 
\\&-Q^2 r_p \left(14 M+3 \lambda  r_p\right)\Big] + q \left(2 Q^2-r_p \left(3 M+\lambda r_p\right)\right) \left(Q^2 r_p \left(r_p(1-\lambda)-7 M\right)+M r_p^2 \left(3 M+\lambda  r_p\right)+3
   Q^4\right)\Bigg) \, Y^{lm}\left(\frac{\pi}{2},0\right) \, , \notag
\\
& H_{lm}^\text{even} = \frac{2 \pi  f_p^2 \left[q \left(r_p \left(3 M+\lambda r_p\right)-2 Q^2\right)-2 \mu  Q f_p r_p u^t\right]}{(\lambda +1) \left(r_p \left(3 M+\lambda r_p\right)-2 Q^2\right]} \, Y^{lm}\left(\frac{\pi}{2},0\right)\, .
\end{align}

\section{Additional master function properties}
\label{sec:AppB}

\subsection{Inverse master function relationships}

Section \ref{sec:master} provides expressions to calculate the fields from the master functions, but it would also be useful to calculate the master functions from the fields. In Regge-Wheeler gauge, the odd-parity master function is constructed according to:
\begin{align}
h^\text{odd}_{lm} &= \frac{1}{\lambda} \left( i \omega r \, h_r^{lm}-2 \, h_t^{lm}+ r \frac{d h_t^{lm}}{dr} + \frac{4 Q}{r} a^\text{odd}_{lm} \right) .
\end{align}
Recall that $a^\text{odd}_{lm}$ has simultaneous roles as field and master function. Similarly, the even-parity master functions are constructed as follows:
\begin{align}
&h^\text{even}_{lm} = \frac{r}{\lambda+1} K^{lm} + \frac{f r^3}{(\lambda+1)\left(r(3 M+\lambda r)-2 Q^2\right)}\left( f\, h_{rr}^{lm} -r\frac{d K^{lm}}{dr} \right) ,
\\
&a^\text{even}_{lm} = \frac{1}{2(\lambda+1)}\left(i\omega r^2 \, a_r^{lm} + r^2 \frac{d a_t^{lm}}{dr} + \frac{Q}{2 f}h_{tt}^{lm} - \frac{Q f r(r(\lambda+2)-M)}{2r(3M+\lambda r)-4 Q^2} h_{rr}^{lm} + \frac{Q f r^3}{r(3 M+\lambda r)-2 Q^2} \frac{d K^{lm}}{dr} \right) .
\end{align}
Through these relationships, it is straightforward to transform back and forth between fields and master functions in Regge-Wheeler gauge.

\subsection{Decoupled master equations}

It was shown by Moncrief~\cite{Monc74a,Monc74b,Monc75} that the coupled homogeneous master equations can be decoupled through a linear transformation of the master functions. Here we demonstrate that procedure in the nonhomogeneous case (and derive appropriate sources for the decoupled equations). The master equations can be written in the following form:
\begin{align}
\left( \frac{d^2}{dr_*^2} + \omega^2 - U_l - W_l \, \mathbf{T}_l \right) \left[\begin{array}{c} h_{lm} \\ a_{lm} \end{array} \right] &= \left[\begin{array}{c} S_{lm} \\ Z_{lm} \end{array} \right] ,
\end{align}
where $U_l$ and $W_l$ are functions of $r$, and $\mathbf{T}_l$ is a matrix of constants. In the odd-parity case these are determined through comparison with Eq.~\eqref{eq:masterOdd},
\begin{align}
&U_l^\text{odd} = \frac{f}{r^2}\left( l(l+1)-\frac{3 M}{r} +\frac{4 Q^2}{r^2} \right),
\\
&W_l^\text{odd} = -\frac{f}{r^3} ,
\\
&\mathbf{T}_l^\text{odd} = \left[\begin{array}{cc} 3 M & 8 Q \\ \lambda Q & -3 M \end{array} \right] ,
\end{align}
and in the even-parity case they are determined through comparison with Eq.~\eqref{eq:masterEven},
\begin{align}
&U_l^\text{even} = f \Big[ r^4 \left(r (3 M+\lambda  r)-2 Q^2\right)^2 \Big]^{-1} \Big(-Q^2 r^2 [39 M^2+32 \lambda  M r+4 (\lambda -1) \lambda r^2] \notag
\\& \qquad\qquad\qquad\qquad +r^3 [9 M^3+9 M^2 r (2 \lambda+1)+3 \lambda  (3 \lambda +2) M r^2+2 \lambda ^2 (\lambda +1)  r^3]+4 Q^4 r (8 M+3 \lambda  r)-8 Q^6 \Big) ,
\\
&W_l^\text{even} = -\frac{f \left(-3 M^2 r+M \left(Q^2+3
   r^2\right)+\lambda  (\lambda +2)
   r^3\right)}{r^2 \left(r (3 M+\lambda  r)-2
   Q^2\right)^2} ,
\\
&\mathbf{T}_l^\text{even} = \left[\begin{array}{cc} 3 M & 8 Q \\ \lambda Q & -3 M \end{array} \right] .
\end{align}
Diagonalizing $\mathbf{T}_l$ (which is the same for even or odd parity) is the key to finding new master functions that conveniently satisfy decoupled equations,
\begin{align}
\mathbf{T}_l = \mathbf{P}_l \, \mathbf{D}_l \, \mathbf{P}^{-1}_l ,
\end{align}
where $\mathbf{D}_l$ is a diagonal matrix consisting of the eigenvalues of $\mathbf{T}_l$, and $\mathbf{P}_l$ is a matrix where each column is an appropriate eigenvector of $\mathbf{T}_l$. Based on this diagonalization, the transformation: 
\begin{align}
\left[\begin{array}{c} h_{lm} \\ a_{lm} \end{array} \right] = \mathbf{P}_l \, \left[\begin{array}{c} R^{(0)}_{lm} \\ R^{(1)}_{lm} \end{array} \right] 
\end{align}
produces decoupled master equations governing new master functions $R^{(0)}_{lm}$ and $R^{(1)}_{lm}$, 
\begin{align}
&\left( \frac{d^2}{dr_*^2} +\omega^2 - V^{(0)}_l \right) R^{(0)}_{lm} = X^{(0)}_{lm} ,
\\
&\left( \frac{d^2}{dr_*^2} +\omega^2 - V^{(1)}_l \right) R^{(1)}_{lm} = X^{(1)}_{lm} .
\end{align}
Note that equivalent decoupled equations are available in~\cite{IshiKoda11} for static spacetimes in an arbitrary number of spatial dimensions. The potentials and source terms for the decoupled master equations are related to the original (coupled) versions,
\begin{align}
& \left[\begin{array}{c} V^{(0)}_l \\ V^{(1)}_l \end{array} \right] = \Big( U_l + W_l \, \mathbf{D}_l \Big) \left[\begin{array}{c} 1 \\ 1 \end{array} \right] ,
\\
& \left[\begin{array}{c} X^{(0)}_{lm} \\ X^{(1)}_{lm} \end{array} \right] =  \mathbf{P}_l^{-1}\, \left[\begin{array}{c} S_{lm} \\ Z_{lm} \end{array} \right].
\end{align}

\section{Low multipole modes}
\label{sec:low-multipole}

\subsection{Even-parity dipole mode}
\label{sec:dipole}
For $l=1$, $m=\pm 1$ the tensor spherical harmonic $Y_{AB}^{1m}$ vanishes, which eliminates Eq.~\eqref{eq:even10} from the field equations and causes $\mathcal{Q}^\sharp_{1m}$ (a source term) and $G^{1m}$ (a mode of the metric perturbation) to vanish. The automatic vanishing of $G^{1m}$ relinquishes 1 degree of gauge freedom, which we fix by enforcing $K^{1m}=0$. It is possible to express $h_{tt}^{1m}$ and $h_{tr}^{1m}$ in terms of $h_{rr}^{1m}$ by forming linear combinations of the nonvanishing even-parity field equations,
\begin{align}
h^{1m}_{tt} &= \frac{r^4f}{Q^2-3Mr}\left( \frac{f}{r^3}\left(r^3 \omega^2 - M \right) h^{1m}_{rr} + \frac{2Q}{r^2}\frac{d a_t^{1m}}{dr} + \frac{4 Q}{r^3}a_t^{1m} + \frac{2i\omega Q}{r^2} a_r^{1m} - \frac{1}{f} \mathcal{Q}^{rr}_{1m}+ i\omega r \mathcal{Q}^{tr}_{1m} -\frac{1}{r} \mathcal{Q}^r_{1m} \right) ,
\\
h^{1m}_{tr} &= i\omega rf h_{rr}^{1m} - r^2 Q^{tr}_{1m} .
\end{align}
With only one unknown piece of the metric perturbation ($h_{rr}^{1m}$) we do not use a gravitational master function for this case, but we do define an electromagnetic master function, $a^\text{even}_{1m}$:
\begin{align}
a_t^{1m} &= f \frac{d a^\text{even}_{1m}}{dr} - \frac{2fQ^2}{r(2Q^2-3Mr)} a^\text{even}_{1m} + \frac{rf^2 Q}{2(2Q^2-3Mr)} h^{1m}_{rr} - \frac{r^3f^2Q}{2(2Q^2-3Mr)} \mathcal{Q}^{tt}_{1m} - \frac{r^2 f}{2}\mathcal{J}^t_{1m},
\\
a_r^{1m} &= -\frac{i\omega}{f} a^\text{even}_{1m} + \frac{i\omega r^2 f Q}{2(2Q^2-3Mr)} h^{1m}_{rr} +\frac{r^2}{2f} \mathcal{J}^r_{1m} .
\end{align}
According to that definition, the field equations reduce to two differential equations governing $a^\text{even}_{1m}$ and $h_{rr}^{1m}$,
\begin{align}
\label{eq:hdipole}
&\qquad \frac{r}{f} \mathcal{Q}^{tt}_{1m} = \frac{d h^{1m}_{rr}}{dr} + \left( \frac{3M}{2Q^2-3Mr}+\frac{5r-4M}{r^2f}-\frac{2}{r} \right) h^{1m}_{rr} - \frac{4Q}{r^3f^2} a^\text{even}_{1m} ,
\\
&\frac{r^2f^2}{2} \frac{d\mathcal{J}^t_{1m}}{dr} - \frac{i\omega r^2}{2} \mathcal{J}^r_{1m}  + \frac{f(Q^2-M r)(3r(r-M)+Q^2)}{r(2Q^2-3Mr)}\mathcal{J}^t_{1m} + \frac{rQ}{2Q^2-3Mr}  \Bigg(\frac{r^2f^3}{2}\frac{d\mathcal{Q}^{tt}_{1m}}{dr}  - \frac{i\omega r^2 f}{2} \mathcal{Q}^{tr}_{1m} + \frac{r}{2}\mathcal{Q}^{rr}_{1m}  \notag
\\& - \frac{f^2 (2MQ^2-6Q^2 r+3Mr^2)}{2(2Q^2-3Mr)}\mathcal{Q}^{tt}_{1m} + \frac{f}{2}\mathcal{Q}^{r}_{1m}  \Bigg) = \frac{d^2 a_{1m}^\text{even}}{dr_*^2} + \left(\omega^2 +\frac{2f(4Q^6-16MQ^4 r+18M^2 Q^2 r^2-9M^2r^4)}{r^4(2Q^2-3Mr)^2} \right) a_{1m}^\text{even} .
\label{eq:adipole}
\end{align}
Note that Eq.~\eqref{eq:adipole} is identical to the general master equation governing $a^\text{even}_{1m}$ (Eq.~\eqref{eq:masterEven} with $l=1$) and is not coupled to $h_{rr}^{1m}$ (although Eq.~\eqref{eq:hdipole} couples $a^\text{even}_{1m}$ to $h_{rr}^{1m}$). See \cite{ZhuOsbu18} for additional details about how we determine the even-parity dipole mode in practice.

\subsection{Odd-parity dipole mode}

For $l=1$, $m=0$ the tensor spherical harmonic $X_{AB}^{10}$ vanishes, which eliminates Eq.~\eqref{eq:odd4} from the field equations and causes $\mathcal{P}_{10}$ (a source term) and $h_2^{10}$ (a mode of the metric perturbation) to vanish. The automatic vanishing of $h_2^{10}$ leaves 1 degree of gauge freedom. Regardless of gauge, linear combinations of Eqs.~\eqref{eq:odd1} and \eqref{eq:odd2}-\eqref{eq:odd3} (and their $r$ derivatives) result in the following decoupled equation governing $a^\text{odd}_{10}$:
\begin{align}
\label{eq:dipole0}
f \mathcal{J}^\flat_{10} - \frac{fQ}{i\omega r^2} \mathcal{P}^r_{10} = \frac{d^2 a_{10}^\text{odd}}{dr_*^2}  + \left( \omega^2 - \frac{f}{r^4}(2 r^2 + 4 Q^2) \right)a_{10}^\text{odd} .
\end{align}
Unlike the even-parity dipole, Eq.~\eqref{eq:dipole0} is not identical to the general master equation governing $a_{lm}^\text{odd}$ (Eq.~\eqref{eq:masterOdd} with $l=1$) because it involves an additional source term. Unfortunately, frequency appears in the denominator of this extra source term, which inhibits consideration of static modes or time-domain strategies (and this issue cannot be avoided by re-expressing the source based on stress-energy conservation). Considering that we have carefully developed our formalism for other modes to avoid exactly this problem, we present here a practical alternative.

Although Eq.~\eqref{eq:dipole0} is gauge invariant, there are specific gauges where $a^{10}_\text{odd}$ can be found via coupled equations without frequency in the denominator. One such gauge involves choosing $h_r^{10}=0$. The following are the nonvanishing field equations (Eqs.~\eqref{eq:odd1} and \eqref{eq:odd2}-\eqref{eq:odd3}) for this gauge:
\begin{align}
&\mathcal{J}^{\flat}_{10} = f\frac{d^2 a_{10}^\text{odd}}{dr^2} +\frac{2(Mr-Q^2)}{r^3}\frac{d a_{10}^\text{odd}}{dr} + \left( \frac{\omega^2}{f} -\frac{2}{r^2} \right) a_{10}^\text{odd} + \frac{Q}{r^2}\frac{d h_t^{10}}{dr} - \frac{2Q}{r^3} h_t^{10} ,
\label{eq:dipole1}
\\
&\mathcal{P}^t_{10} = \frac{d^2h_t^{10}}{dr^2} - \frac{2}{r^2} h_t^{10} + \frac{4Q}{r^2}\frac{da^\text{odd}_{10}}{dr} ,
\label{eq:dipole2}
\\
&\mathcal{P}^r_{10} = i\omega \frac{dh_t^{10}}{dr}-\frac{2 i \omega}{r} h_t^{10} + \frac{4i\omega Q}{r^2} a^\text{odd}_{10} .
\label{eq:dipole3}
\end{align}
Equations~\eqref{eq:dipole2} and \eqref{eq:dipole3} are redundant because linear combinations of Eq.~\eqref{eq:dipole3} and its $r$ derivative are identical to Eq.~\eqref{eq:dipole2} (after appropriate application of stress-energy conservation). Therefore, we believe that Eqs.~\eqref{eq:dipole1} and \eqref{eq:dipole2} form a coupled system that is straightforward to solve (perhaps after adjustment for desirable numerical properties) for $a^\text{odd}_{10}$ and $h_t^{10}$ regardless of whether static modes or the time-domain are involved. Although, we should note that we have not actually calculated numerical solutions for the odd-parity dipole case because it is nonradiative for point particles in circular motion. 

\subsection{Monopole mode}

For $l=0$, $m=0$ the tensor and vector spherical harmonics $Y_{AB}^{00}$ and $Y_{A}^{00}$ vanish, which eliminates Eqs.~\eqref{eq:even3},  \eqref{eq:even7}-\eqref{eq:even8}, and \eqref{eq:even10} from the field equations (and $\mathcal{J}^\sharp_{00}$, $\mathcal{Q}^t_{00}$, $\mathcal{Q}^r_{00}$, $\mathcal{Q}^\sharp_{00}$, $a^{00}_\sharp$, $j_t^{00}$, $j_r^{00}$, and $G^{00}$ also vanish). Three degrees of gauge freedom remain after the automatic vanishing of aforementioned fields. We fix the gauge by enforcing $a^{00}_r =  h^{00}_{tr} =  K^{00} = 0$. The following are the nonvanishing field equations (Eqs.~\eqref{eq:even1}-\eqref{eq:even2}, \eqref{eq:even4}-\eqref{eq:even6}, and \eqref{eq:even9}) for this gauge:
\begin{align}
&\mathcal{J}^t_{00} = \frac{d^2 a_t^{00}}{dr^2} + \frac{2}{r}\frac{d a_t^{00}}{dr} +\frac{Q}{2r^2f}\frac{d h_{tt}^{00}}{dr} - \frac{fQ}{2r^2} \frac{d h_{rr}^{00}}{dr} +\frac{Q(Q^2-M r)}{r^5 f^2}  \left( h_{tt}^{00} + f^2 h_{rr}^{00} \right) ,
\label{eq:mono1}
\\
&\mathcal{J}^r_{00} = i \omega \frac{da_t^{00}}{dr} + \frac{i\omega Q}{2 r^2f} h_{tt}^{00} - \frac{i\omega f Q}{2 r^2} h_{rr}^{00} ,
\label{eq:mono2}
\\
&\mathcal{Q}^{tt}_{00} =\frac{f}{r} \frac{dh_{rr}^{00}}{dr} - \frac{2Q^2-r(2M+r)}{r^4} h_{rr}^{00} -\frac{Q^2}{r^4f^2} h_{tt}^{00} -\frac{2Q}{r^2 f}\frac{da_{t}^{00}}{dr} ,
\label{eq:mono3}
\\
&\mathcal{Q}^{tr}_{00} = \frac{i\omega f}{r} h_{rr}^{00} ,
\label{eq:mono4}
\\
&\mathcal{Q}^{rr}_{00} = - \frac{f}{r}\frac{dh_{tt}^{00}}{dr} - \frac{f^2}{r^2}h_{rr}^{00} - \frac{Q^2-2Mr}{r^4} h_{tt}^{00} +\frac{2fQ}{r^2}\frac{da_t^{00}}{dr} ,
\label{eq:mono5}
\\
&\mathcal{Q}^\flat_{00} = - \frac{d^2h_{tt}^{00}}{dr^2} - \left( \frac{2}{r} -\frac{r-M}{r^2f}\right)\frac{dh_{tt}^{00}}{dr}-\frac{f(r-M)}{r^2}\frac{dh_{rr}^{00}}{dr} + \frac{2(r-M)(Q^2-Mr)}{f^2 r^5}\left( h_{tt}^{00} +f^2 h_{rr}^{00} \right) + \omega^2 h_{rr}^{00} - \frac{4Q}{r^2}\frac{da_t^{00}}{dr} .
\label{eq:mono6}
\end{align}
Similar to Eqs.~\eqref{eq:dipole2} and \eqref{eq:dipole3}, many of the monopole equations (like Eqs.~\eqref{eq:mono1} and \eqref{eq:mono2}) are redundant. We propose an approach where first the $r$ derivative of $h_{rr}^{00}$ is eliminated from the field equations via Eq.~\eqref{eq:mono3},
\begin{align}
\frac{d h_{rr}^{00}}{dr} &= \frac{2Q^2-r(2M+r)}{r^3 f} h_{rr}^{00} + \frac{Q^2}{r^3 f^3}h_{tt}^{00} + \frac{2 Q}{r f^2}\frac{d a_t^{00}}{dr} + \frac{r}{f} \mathcal{Q}^{tt}_{00} ,
\end{align}
and then $h_{rr}^{00}$ is subsequently eliminated from the field equations via Eq.~\eqref{eq:mono5},
\begin{align}
h_{rr}^{00} &= -\frac{r}{f}\frac{d h_{tt}^{00}}{dr} - \frac{Q^2-2Mr}{r^2 f^2} + \frac{2Q}{f}\frac{d a_t^{00}}{dr} - \frac{r^2}{f^2} \mathcal{Q}^{rr}_{00} .
\end{align}
The remaining unknowns, $a_t^{00}$ and $h_{tt}^{00}$, are governed by a reduced system with refactored Eqs.~\eqref{eq:mono1} and \eqref{eq:mono6},
\begin{align}
&\mathcal{J}^t_{00} + \frac{Q}{2r}\mathcal{Q}^{tt}_{00} + \frac{Q}{2 r f^2} \mathcal{Q}^{rr}_{00} = \frac{d^2 a_t^{00}}{dr^2}+\frac{2}{r}\frac{d a_t^{00}}{dr} ,
\\
&\mathcal{Q}^\flat_{00} + \frac{r-M}{r}\mathcal{Q}^{tt}_{00}  + \frac{r-M+\omega^2 r^3}{r f^2}\mathcal{Q}^{rr}_{00} = -\frac{d^2 h_{tt}^{00}}{dr^2}-\frac{2}{r}\frac{d h_{tt}^{00}}{dr} +\frac{4Q}{r^2}\frac{d a_t^{00}}{dr} \notag
\\&\qquad\qquad\qquad\qquad\qquad\qquad\qquad\qquad\qquad\qquad - \omega^2\left(\frac{Q^2 - 2 M r}{r^2 f^2}h_{tt}^{00} +\frac{r}{f} \frac{d h_{tt}^{00}}{dr} +\frac{2 Q}{f} \frac{d a_t^{00}}{dr} \right) .
\end{align}
Similar to the odd-parity dipole case, we have not calculated numerical solutions for the monopole mode because it is nonradiative, although we predict there may be subtleties related to consistency between the perturbed mass-energy and properties of the source.

\end{appendix}

\end{widetext}

\bibliography{charges}

\begin{thebibliography}{50}%
\makeatletter
\providecommand \@ifxundefined [1]{%
 \@ifx{#1\undefined}
}%
\providecommand \@ifnum [1]{%
 \ifnum #1\expandafter \@firstoftwo
 \else \expandafter \@secondoftwo
 \fi
}%
\providecommand \@ifx [1]{%
 \ifx #1\expandafter \@firstoftwo
 \else \expandafter \@secondoftwo
 \fi
}%
\providecommand \natexlab [1]{#1}%
\providecommand \enquote  [1]{``#1''}%
\providecommand \bibnamefont  [1]{#1}%
\providecommand \bibfnamefont [1]{#1}%
\providecommand \citenamefont [1]{#1}%
\providecommand \href@noop [0]{\@secondoftwo}%
\providecommand \href [0]{\begingroup \@sanitize@url \@href}%
\providecommand \@href[1]{\@@startlink{#1}\@@href}%
\providecommand \@@href[1]{\endgroup#1\@@endlink}%
\providecommand \@sanitize@url [0]{\catcode `\\12\catcode `\$12\catcode
  `\&12\catcode `\#12\catcode `\^12\catcode `\_12\catcode `\%12\relax}%
\providecommand \@@startlink[1]{}%
\providecommand \@@endlink[0]{}%
\providecommand \url  [0]{\begingroup\@sanitize@url \@url }%
\providecommand \@url [1]{\endgroup\@href {#1}{\urlprefix }}%
\providecommand \urlprefix  [0]{URL }%
\providecommand \Eprint [0]{\href }%
\providecommand \doibase [0]{http://dx.doi.org/}%
\providecommand \selectlanguage [0]{\@gobble}%
\providecommand \bibinfo  [0]{\@secondoftwo}%
\providecommand \bibfield  [0]{\@secondoftwo}%
\providecommand \translation [1]{[#1]}%
\providecommand \BibitemOpen [0]{}%
\providecommand \bibitemStop [0]{}%
\providecommand \bibitemNoStop [0]{.\EOS\space}%
\providecommand \EOS [0]{\spacefactor3000\relax}%
\providecommand \BibitemShut  [1]{\csname bibitem#1\endcsname}%
\let\auto@bib@innerbib\@empty
\bibitem [{\citenamefont {Rees}(1984)}]{Rees84}%
  \BibitemOpen
  \bibfield  {author} {\bibinfo {author} {\bibfnamefont {M.~J.}\ \bibnamefont
  {Rees}},\ }\href {\doibase 10.1146/annurev.aa.22.090184.002351} {\bibfield
  {journal} {\bibinfo  {journal} {Annu. Rev. of Astron. and Astrophys.}\
  }\textbf {\bibinfo {volume} {22}},\ \bibinfo {pages} {471} (\bibinfo {year}
  {1984})}\BibitemShut {NoStop}%
\bibitem [{\citenamefont {Abbott}\ \emph
  {et~al.}(2016{\natexlab{a}})\citenamefont {Abbott} \emph {et~al.}}]{LIGO1}%
  \BibitemOpen
  \bibfield  {author} {\bibinfo {author} {\bibfnamefont {B.~P.}\ \bibnamefont
  {Abbott}} \emph {et~al.} (\bibinfo {collaboration} {LIGO Scientific and Virgo
  Collaborations}),\ }\href {\doibase 10.1103/PhysRevLett.116.061102}
  {\bibfield  {journal} {\bibinfo  {journal} {Phys. Rev. Lett.}\ }\textbf
  {\bibinfo {volume} {116}},\ \bibinfo {pages} {061102} (\bibinfo {year}
  {2016}{\natexlab{a}})}\BibitemShut {NoStop}%
\bibitem [{\citenamefont {Abbott}\ \emph
  {et~al.}(2016{\natexlab{b}})\citenamefont {Abbott} \emph {et~al.}}]{LIGO2}%
  \BibitemOpen
  \bibfield  {author} {\bibinfo {author} {\bibfnamefont {B.~P.}\ \bibnamefont
  {Abbott}} \emph {et~al.} (\bibinfo {collaboration} {LIGO Scientific and Virgo
  Collaborations}),\ }\href {\doibase 10.1103/PhysRevLett.116.241103}
  {\bibfield  {journal} {\bibinfo  {journal} {Phys. Rev. Lett.}\ }\textbf
  {\bibinfo {volume} {116}},\ \bibinfo {pages} {241103} (\bibinfo {year}
  {2016}{\natexlab{b}})}\BibitemShut {NoStop}%
\bibitem [{\citenamefont {Abbott}\ \emph
  {et~al.}(2017{\natexlab{a}})\citenamefont {Abbott} \emph {et~al.}}]{LIGO3}%
  \BibitemOpen
  \bibfield  {author} {\bibinfo {author} {\bibfnamefont {B.~P.}\ \bibnamefont
  {Abbott}} \emph {et~al.} (\bibinfo {collaboration} {LIGO Scientific and Virgo
  Collaborations}),\ }\href {\doibase 10.1103/PhysRevLett.118.221101}
  {\bibfield  {journal} {\bibinfo  {journal} {Phys. Rev. Lett.}\ }\textbf
  {\bibinfo {volume} {118}},\ \bibinfo {pages} {221101} (\bibinfo {year}
  {2017}{\natexlab{a}})}\BibitemShut {NoStop}%
\bibitem [{\citenamefont {Abbott}\ \emph
  {et~al.}(2017{\natexlab{b}})\citenamefont {Abbott} \emph {et~al.}}]{LIGO4}%
  \BibitemOpen
  \bibfield  {author} {\bibinfo {author} {\bibfnamefont {B.~P.}\ \bibnamefont
  {Abbott}} \emph {et~al.} (\bibinfo {collaboration} {LIGO Scientific and Virgo
  Collaborations}),\ }\href {\doibase 10.3847/2041-8213/aa9f0c} {\bibfield
  {journal} {\bibinfo  {journal} {Astrophys. J. Lett.}\ }\textbf {\bibinfo
  {volume} {851}},\ \bibinfo {pages} {L35} (\bibinfo {year}
  {2017}{\natexlab{b}})}\BibitemShut {NoStop}%
\bibitem [{\citenamefont {Abbott}\ \emph
  {et~al.}(2017{\natexlab{c}})\citenamefont {Abbott} \emph {et~al.}}]{LIGO5}%
  \BibitemOpen
  \bibfield  {author} {\bibinfo {author} {\bibfnamefont {B.~P.}\ \bibnamefont
  {Abbott}} \emph {et~al.} (\bibinfo {collaboration} {LIGO Scientific and Virgo
  Collaborations}),\ }\href {\doibase 10.1103/PhysRevLett.119.141101}
  {\bibfield  {journal} {\bibinfo  {journal} {Phys. Rev. Lett.}\ }\textbf
  {\bibinfo {volume} {119}},\ \bibinfo {pages} {141101} (\bibinfo {year}
  {2017}{\natexlab{c}})}\BibitemShut {NoStop}%
\bibitem [{\citenamefont {Abbott}\ \emph
  {et~al.}(2020{\natexlab{a}})\citenamefont {Abbott} \emph {et~al.}}]{LIGO7}%
  \BibitemOpen
  \bibfield  {author} {\bibinfo {author} {\bibfnamefont {B.~P.}\ \bibnamefont
  {Abbott}} \emph {et~al.} (\bibinfo {collaboration} {LIGO Scientific and Virgo
  Collaborations}),\ }\href {\doibase 10.3847/2041-8213/ab75f5} {\bibfield
  {journal} {\bibinfo  {journal} {Astrophys. J.}\ }\textbf {\bibinfo {volume}
  {892}},\ \bibinfo {pages} {L3} (\bibinfo {year}
  {2020}{\natexlab{a}})}\BibitemShut {NoStop}%
\bibitem [{\citenamefont {Abbott}\ \emph
  {et~al.}(2020{\natexlab{b}})\citenamefont {Abbott} \emph {et~al.}}]{LIGO8}%
  \BibitemOpen
  \bibfield  {author} {\bibinfo {author} {\bibfnamefont {B.~P.}\ \bibnamefont
  {Abbott}} \emph {et~al.} (\bibinfo {collaboration} {LIGO Scientific and Virgo
  Collaborations}),\ }\href {\doibase 10.3847/2041-8213/ab960f} {\bibfield
  {journal} {\bibinfo  {journal} {Astrophys. J.}\ }\textbf {\bibinfo {volume}
  {896}},\ \bibinfo {pages} {L44} (\bibinfo {year}
  {2020}{\natexlab{b}})}\BibitemShut {NoStop}%
\bibitem [{\citenamefont {Narayan}(2005)}]{Nara05}%
  \BibitemOpen
  \bibfield  {author} {\bibinfo {author} {\bibfnamefont {R.}~\bibnamefont
  {Narayan}},\ }\href {\doibase 10.1088/1367-2630/7/1/199} {\bibfield
  {journal} {\bibinfo  {journal} {New J. Phys.}\ }\textbf {\bibinfo {volume}
  {7}},\ \bibinfo {pages} {199} (\bibinfo {year} {2005})}\BibitemShut {NoStop}%
\bibitem [{\citenamefont {Wald}(1974)}]{Wald74}%
  \BibitemOpen
  \bibfield  {author} {\bibinfo {author} {\bibfnamefont {R.~M.}\ \bibnamefont
  {Wald}},\ }\href {\doibase 10.1103/PhysRevD.10.1680} {\bibfield  {journal}
  {\bibinfo  {journal} {Phys. Rev. D}\ }\textbf {\bibinfo {volume} {10}},\
  \bibinfo {pages} {1680} (\bibinfo {year} {1974})}\BibitemShut {NoStop}%
\bibitem [{\citenamefont {Ruffini}(1977)}]{Ruff77}%
  \BibitemOpen
  \bibfield  {author} {\bibinfo {author} {\bibfnamefont {R.}~\bibnamefont
  {Ruffini}},\ }\enquote {\bibinfo {title} {in $\text{P}$roceedings of 1st
  $\text{M}$arcel $\text{G}$rossmann $\text{M}$eeting},}\ \ (\bibinfo
  {publisher} {North-Holland},\ \bibinfo {address} {Amsterdam, New York},\
  \bibinfo {year} {1977})\ p.\ \bibinfo {pages} {349}\BibitemShut {NoStop}%
\bibitem [{\citenamefont {{Punsly}}(1998)}]{Puns98}%
  \BibitemOpen
  \bibfield  {author} {\bibinfo {author} {\bibfnamefont {B.}~\bibnamefont
  {{Punsly}}},\ }\href {\doibase 10.1086/305561} {\bibfield  {journal}
  {\bibinfo  {journal} {\apj}\ }\textbf {\bibinfo {volume} {498}},\ \bibinfo
  {pages} {640} (\bibinfo {year} {1998})}\BibitemShut {NoStop}%
\bibitem [{\citenamefont {De~Rujula}\ \emph {et~al.}(1990)\citenamefont
  {De~Rujula}, \citenamefont {Glashow},\ and\ \citenamefont
  {Sarid}}]{RujuETC90}%
  \BibitemOpen
  \bibfield  {author} {\bibinfo {author} {\bibfnamefont {A.}~\bibnamefont
  {De~Rujula}}, \bibinfo {author} {\bibfnamefont {S.}~\bibnamefont {Glashow}},
  \ and\ \bibinfo {author} {\bibfnamefont {U.}~\bibnamefont {Sarid}},\ }\href
  {\doibase 10.1016/0550-3213(90)90227-5} {\bibfield  {journal} {\bibinfo
  {journal} {Nucl. Phys.}\ }\textbf {\bibinfo {volume} {B333}},\ \bibinfo
  {pages} {173} (\bibinfo {year} {1990})}\BibitemShut {NoStop}%
\bibitem [{\citenamefont {Cardoso}\ \emph {et~al.}(2016)\citenamefont
  {Cardoso}, \citenamefont {Macedo}, \citenamefont {Pani},\ and\ \citenamefont
  {Ferrari}}]{CardETC16}%
  \BibitemOpen
  \bibfield  {author} {\bibinfo {author} {\bibfnamefont {V.}~\bibnamefont
  {Cardoso}}, \bibinfo {author} {\bibfnamefont {C.~F.~B.}\ \bibnamefont
  {Macedo}}, \bibinfo {author} {\bibfnamefont {P.}~\bibnamefont {Pani}}, \ and\
  \bibinfo {author} {\bibfnamefont {V.}~\bibnamefont {Ferrari}},\ }\href
  {\doibase 10.1088/1475-7516/2016/05/054} {\bibfield  {journal} {\bibinfo
  {journal} {J. Cosmol. Astropart. Phys.}\ }\textbf {\bibinfo {volume}
  {2016}},\ \bibinfo {pages} {054} (\bibinfo {year} {2016})}\BibitemShut
  {NoStop}%
\bibitem [{\citenamefont {{Zerilli}}(1974)}]{Zeri74}%
  \BibitemOpen
  \bibfield  {author} {\bibinfo {author} {\bibfnamefont {F.~J.}\ \bibnamefont
  {{Zerilli}}},\ }\href {\doibase 10.1103/PhysRevD.9.860} {\bibfield  {journal}
  {\bibinfo  {journal} {Phys. Rev. D}\ }\textbf {\bibinfo {volume} {9}},\
  \bibinfo {pages} {860} (\bibinfo {year} {1974})}\BibitemShut {NoStop}%
\bibitem [{\citenamefont {Moncrief}(1974{\natexlab{a}})}]{Monc74a}%
  \BibitemOpen
  \bibfield  {author} {\bibinfo {author} {\bibfnamefont {V.}~\bibnamefont
  {Moncrief}},\ }\href {\doibase 10.1103/PhysRevD.9.2707} {\bibfield  {journal}
  {\bibinfo  {journal} {Phys. Rev. D}\ }\textbf {\bibinfo {volume} {9}},\
  \bibinfo {pages} {2707} (\bibinfo {year} {1974}{\natexlab{a}})}\BibitemShut
  {NoStop}%
\bibitem [{\citenamefont {Moncrief}(1974{\natexlab{b}})}]{Monc74b}%
  \BibitemOpen
  \bibfield  {author} {\bibinfo {author} {\bibfnamefont {V.}~\bibnamefont
  {Moncrief}},\ }\href {\doibase 10.1103/PhysRevD.10.1057} {\bibfield
  {journal} {\bibinfo  {journal} {Phys. Rev. D}\ }\textbf {\bibinfo {volume}
  {10}},\ \bibinfo {pages} {1057} (\bibinfo {year}
  {1974}{\natexlab{b}})}\BibitemShut {NoStop}%
\bibitem [{\citenamefont {Moncrief}(1975)}]{Monc75}%
  \BibitemOpen
  \bibfield  {author} {\bibinfo {author} {\bibfnamefont {V.}~\bibnamefont
  {Moncrief}},\ }\href {\doibase 10.1103/PhysRevD.12.1526} {\bibfield
  {journal} {\bibinfo  {journal} {Phys. Rev. D}\ }\textbf {\bibinfo {volume}
  {12}},\ \bibinfo {pages} {1526} (\bibinfo {year} {1975})}\BibitemShut
  {NoStop}%
\bibitem [{\citenamefont {Zhu}\ and\ \citenamefont {Osburn}(2018)}]{ZhuOsbu18}%
  \BibitemOpen
  \bibfield  {author} {\bibinfo {author} {\bibfnamefont {R.}~\bibnamefont
  {Zhu}}\ and\ \bibinfo {author} {\bibfnamefont {T.}~\bibnamefont {Osburn}},\
  }\href {\doibase 10.1103/PhysRevD.97.104058} {\bibfield  {journal} {\bibinfo
  {journal} {Phys. Rev. D}\ }\textbf {\bibinfo {volume} {97}},\ \bibinfo
  {pages} {104058} (\bibinfo {year} {2018})}\BibitemShut {NoStop}%
\bibitem [{\citenamefont {{Cohen}}\ \emph {et~al.}(1975)\citenamefont
  {{Cohen}}, \citenamefont {{Kegeles}},\ and\ \citenamefont
  {{Rosenblum}}}]{Cohe75}%
  \BibitemOpen
  \bibfield  {author} {\bibinfo {author} {\bibfnamefont {J.~M.}\ \bibnamefont
  {{Cohen}}}, \bibinfo {author} {\bibfnamefont {L.~S.}\ \bibnamefont
  {{Kegeles}}}, \ and\ \bibinfo {author} {\bibfnamefont {A.}~\bibnamefont
  {{Rosenblum}}},\ }\href {\doibase 10.1086/153944} {\bibfield  {journal}
  {\bibinfo  {journal} {Astrophys. J.}\ }\textbf {\bibinfo {volume} {201}},\
  \bibinfo {pages} {783} (\bibinfo {year} {1975})}\BibitemShut {NoStop}%
\bibitem [{\citenamefont {Johnston}\ \emph {et~al.}(1973)\citenamefont
  {Johnston}, \citenamefont {Ruffini},\ and\ \citenamefont {Zerilli}}]{John73}%
  \BibitemOpen
  \bibfield  {author} {\bibinfo {author} {\bibfnamefont {M.}~\bibnamefont
  {Johnston}}, \bibinfo {author} {\bibfnamefont {R.}~\bibnamefont {Ruffini}}, \
  and\ \bibinfo {author} {\bibfnamefont {F.}~\bibnamefont {Zerilli}},\ }\href
  {\doibase 10.1103/PhysRevLett.31.1317} {\bibfield  {journal} {\bibinfo
  {journal} {Phys. Rev. Lett.}\ }\textbf {\bibinfo {volume} {31}},\ \bibinfo
  {pages} {1317} (\bibinfo {year} {1973})}\BibitemShut {NoStop}%
\bibitem [{\citenamefont {Johnston}\ \emph {et~al.}(1974)\citenamefont
  {Johnston}, \citenamefont {Ruffini},\ and\ \citenamefont {Zerilli}}]{John74}%
  \BibitemOpen
  \bibfield  {author} {\bibinfo {author} {\bibfnamefont {M.}~\bibnamefont
  {Johnston}}, \bibinfo {author} {\bibfnamefont {R.}~\bibnamefont {Ruffini}}, \
  and\ \bibinfo {author} {\bibfnamefont {F.}~\bibnamefont {Zerilli}},\ }\href
  {\doibase 10.1016/0370-2693(74)90505-X} {\bibfield  {journal} {\bibinfo
  {journal} {Phys. Lett.}\ }\textbf {\bibinfo {volume} {B49}},\ \bibinfo
  {pages} {185 } (\bibinfo {year} {1974})}\BibitemShut {NoStop}%
\bibitem [{\citenamefont {Zilh{\~a}o}\ \emph {et~al.}(2012)\citenamefont
  {Zilh{\~a}o}, \citenamefont {Cardoso}, \citenamefont {Herdeiro},
  \citenamefont {Lehner},\ and\ \citenamefont {Sperhake}}]{Zilh12}%
  \BibitemOpen
  \bibfield  {author} {\bibinfo {author} {\bibfnamefont {M.}~\bibnamefont
  {Zilh{\~a}o}}, \bibinfo {author} {\bibfnamefont {V.}~\bibnamefont {Cardoso}},
  \bibinfo {author} {\bibfnamefont {C.}~\bibnamefont {Herdeiro}}, \bibinfo
  {author} {\bibfnamefont {L.}~\bibnamefont {Lehner}}, \ and\ \bibinfo {author}
  {\bibfnamefont {U.}~\bibnamefont {Sperhake}},\ }\href {\doibase
  10.1103/PhysRevD.85.124062} {\bibfield  {journal} {\bibinfo  {journal} {Phys.
  Rev. D}\ }\textbf {\bibinfo {volume} {85}},\ \bibinfo {pages} {124062}
  (\bibinfo {year} {2012})}\BibitemShut {NoStop}%
\bibitem [{\citenamefont {Zilh{\~a}o}\ \emph {et~al.}(2014)\citenamefont
  {Zilh{\~a}o}, \citenamefont {Cardoso}, \citenamefont {Herdeiro},
  \citenamefont {Lehner},\ and\ \citenamefont {Sperhake}}]{Zilh14}%
  \BibitemOpen
  \bibfield  {author} {\bibinfo {author} {\bibfnamefont {M.}~\bibnamefont
  {Zilh{\~a}o}}, \bibinfo {author} {\bibfnamefont {V.}~\bibnamefont {Cardoso}},
  \bibinfo {author} {\bibfnamefont {C.}~\bibnamefont {Herdeiro}}, \bibinfo
  {author} {\bibfnamefont {L.}~\bibnamefont {Lehner}}, \ and\ \bibinfo {author}
  {\bibfnamefont {U.}~\bibnamefont {Sperhake}},\ }\href {\doibase
  10.1103/PhysRevD.89.044008} {\bibfield  {journal} {\bibinfo  {journal} {Phys.
  Rev. D}\ }\textbf {\bibinfo {volume} {89}},\ \bibinfo {pages} {044008}
  (\bibinfo {year} {2014})}\BibitemShut {NoStop}%
\bibitem [{\citenamefont {Abbott}\ \emph
  {et~al.}(2017{\natexlab{d}})\citenamefont {Abbott} \emph {et~al.}}]{LIGO6}%
  \BibitemOpen
  \bibfield  {author} {\bibinfo {author} {\bibfnamefont {B.~P.}\ \bibnamefont
  {Abbott}} \emph {et~al.} (\bibinfo {collaboration} {LIGO Scientific and Virgo
  Collaborations}),\ }\href {\doibase 10.1103/PhysRevLett.119.161101}
  {\bibfield  {journal} {\bibinfo  {journal} {Phys. Rev. Lett.}\ }\textbf
  {\bibinfo {volume} {119}},\ \bibinfo {pages} {161101} (\bibinfo {year}
  {2017}{\natexlab{d}})}\BibitemShut {NoStop}%
\bibitem [{\citenamefont {{Amaro-Seoane}}\ \emph {et~al.}(2017)\citenamefont
  {{Amaro-Seoane}} \emph {et~al.}}]{LISA}%
  \BibitemOpen
  \bibfield  {author} {\bibinfo {author} {\bibfnamefont {P.}~\bibnamefont
  {{Amaro-Seoane}}} \emph {et~al.},\ }\href@noop {} {\  (\bibinfo {year}
  {2017})},\ \Eprint {http://arxiv.org/abs/1702.00786} {arXiv:1702.00786}
  \BibitemShut {NoStop}%
\bibitem [{\citenamefont {Harry}\ and\ \citenamefont {$\text{LIGO Scientific
  Collaboration}$}(2010)}]{aLIGO}%
  \BibitemOpen
  \bibfield  {author} {\bibinfo {author} {\bibfnamefont {G.~M.}\ \bibnamefont
  {Harry}}\ and\ \bibinfo {author} {\bibnamefont {$\text{LIGO Scientific
  Collaboration}$}},\ }\href {\doibase 10.1088/0264-9381/27/8/084006}
  {\bibfield  {journal} {\bibinfo  {journal} {Classical Quantum Gravity}\
  }\textbf {\bibinfo {volume} {27}},\ \bibinfo {pages} {084006} (\bibinfo
  {year} {2010})}\BibitemShut {NoStop}%
\bibitem [{\citenamefont {{Acernese}}\ \emph {et~al.}(2015)\citenamefont
  {{Acernese}} \emph {et~al.}}]{aVIRGO}%
  \BibitemOpen
  \bibfield  {author} {\bibinfo {author} {\bibfnamefont {F.}~\bibnamefont
  {{Acernese}}} \emph {et~al.} (\bibinfo {collaboration} {Virgo
  Collaboration}),\ }\href {\doibase 10.1088/0264-9381/32/2/024001} {\bibfield
  {journal} {\bibinfo  {journal} {Classical Quantum Gravity}\ }\textbf
  {\bibinfo {volume} {32}},\ \bibinfo {eid} {024001} (\bibinfo {year}
  {2015})}\BibitemShut {NoStop}%
\bibitem [{\citenamefont {Mino}\ \emph {et~al.}(1997)\citenamefont {Mino},
  \citenamefont {Sasaki},\ and\ \citenamefont {Tanaka}}]{MinoSasaTana97}%
  \BibitemOpen
  \bibfield  {author} {\bibinfo {author} {\bibfnamefont {Y.}~\bibnamefont
  {Mino}}, \bibinfo {author} {\bibfnamefont {M.}~\bibnamefont {Sasaki}}, \ and\
  \bibinfo {author} {\bibfnamefont {T.}~\bibnamefont {Tanaka}},\ }\href
  {10.1103/PhysRevD.55.3457} {\bibfield  {journal} {\bibinfo  {journal} {Phys.
  Rev. D}\ }\textbf {\bibinfo {volume} {55}},\ \bibinfo {pages} {3457}
  (\bibinfo {year} {1997})}\BibitemShut {NoStop}%
\bibitem [{\citenamefont {Quinn}\ and\ \citenamefont
  {Wald}(1997)}]{QuinWald97}%
  \BibitemOpen
  \bibfield  {author} {\bibinfo {author} {\bibfnamefont {T.~C.}\ \bibnamefont
  {Quinn}}\ and\ \bibinfo {author} {\bibfnamefont {R.~M.}\ \bibnamefont
  {Wald}},\ }\href {\doibase 10.1103/PhysRevD.56.3381} {\bibfield  {journal}
  {\bibinfo  {journal} {Phys. Rev. D}\ }\textbf {\bibinfo {volume} {56}},\
  \bibinfo {pages} {3381} (\bibinfo {year} {1997})}\BibitemShut {NoStop}%
\bibitem [{\citenamefont {Poisson}\ \emph {et~al.}(2011)\citenamefont
  {Poisson}, \citenamefont {Pound},\ and\ \citenamefont {Vega}}]{Pois11}%
  \BibitemOpen
  \bibfield  {author} {\bibinfo {author} {\bibfnamefont {E.}~\bibnamefont
  {Poisson}}, \bibinfo {author} {\bibfnamefont {A.}~\bibnamefont {Pound}}, \
  and\ \bibinfo {author} {\bibfnamefont {I.}~\bibnamefont {Vega}},\ }\href
  {\doibase 10.12942/lrr-2011-7} {\bibfield  {journal} {\bibinfo  {journal}
  {Living Rev. Relativity}\ }\textbf {\bibinfo {volume} {14}},\ \bibinfo
  {pages} {7} (\bibinfo {year} {2011})}\BibitemShut {NoStop}%
\bibitem [{\citenamefont {Barack}(2000)}]{Bara00}%
  \BibitemOpen
  \bibfield  {author} {\bibinfo {author} {\bibfnamefont {L.}~\bibnamefont
  {Barack}},\ }\href {\doibase 10.1103/PhysRevD.62.084027} {\bibfield
  {journal} {\bibinfo  {journal} {Phys. Rev. D}\ }\textbf {\bibinfo {volume}
  {62}},\ \bibinfo {pages} {084027} (\bibinfo {year} {2000})}\BibitemShut
  {NoStop}%
\bibitem [{\citenamefont {Burko}\ and\ \citenamefont {Liu}(2001)}]{Burk01}%
  \BibitemOpen
  \bibfield  {author} {\bibinfo {author} {\bibfnamefont {L.~M.}\ \bibnamefont
  {Burko}}\ and\ \bibinfo {author} {\bibfnamefont {Y.~T.}\ \bibnamefont
  {Liu}},\ }\href {\doibase 10.1103/PhysRevD.64.024006} {\bibfield  {journal}
  {\bibinfo  {journal} {Phys. Rev. D}\ }\textbf {\bibinfo {volume} {64}},\
  \bibinfo {pages} {024006} (\bibinfo {year} {2001})}\BibitemShut {NoStop}%
\bibitem [{\citenamefont {Zimmerman}\ and\ \citenamefont
  {Poisson}(2014)}]{ZimmPois14}%
  \BibitemOpen
  \bibfield  {author} {\bibinfo {author} {\bibfnamefont {P.}~\bibnamefont
  {Zimmerman}}\ and\ \bibinfo {author} {\bibfnamefont {E.}~\bibnamefont
  {Poisson}},\ }\href {\doibase 10.1103/PhysRevD.90.084030} {\bibfield
  {journal} {\bibinfo  {journal} {Phys. Rev. D}\ }\textbf {\bibinfo {volume}
  {90}},\ \bibinfo {pages} {084030} (\bibinfo {year} {2014})}\BibitemShut
  {NoStop}%
\bibitem [{\citenamefont {Linz}\ \emph {et~al.}(2014)\citenamefont {Linz},
  \citenamefont {Friedman},\ and\ \citenamefont {Wiseman}}]{Linz14}%
  \BibitemOpen
  \bibfield  {author} {\bibinfo {author} {\bibfnamefont {T.~M.}\ \bibnamefont
  {Linz}}, \bibinfo {author} {\bibfnamefont {J.~L.}\ \bibnamefont {Friedman}},
  \ and\ \bibinfo {author} {\bibfnamefont {A.~G.}\ \bibnamefont {Wiseman}},\
  }\href {\doibase 10.1103/PhysRevD.90.084031} {\bibfield  {journal} {\bibinfo
  {journal} {Phys. Rev. D}\ }\textbf {\bibinfo {volume} {90}},\ \bibinfo
  {pages} {084031} (\bibinfo {year} {2014})}\BibitemShut {NoStop}%
\bibitem [{\citenamefont {Zimmerman}(2015)}]{Zimm15}%
  \BibitemOpen
  \bibfield  {author} {\bibinfo {author} {\bibfnamefont {P.}~\bibnamefont
  {Zimmerman}},\ }\href {\doibase 10.1103/PhysRevD.92.064040} {\bibfield
  {journal} {\bibinfo  {journal} {Phys. Rev. D}\ }\textbf {\bibinfo {volume}
  {92}},\ \bibinfo {pages} {064040} (\bibinfo {year} {2015})}\BibitemShut
  {NoStop}%
\bibitem [{\citenamefont {Castillo}\ \emph {et~al.}(2018)\citenamefont
  {Castillo}, \citenamefont {Vega},\ and\ \citenamefont
  {Wardell}}]{CastVegaWard18}%
  \BibitemOpen
  \bibfield  {author} {\bibinfo {author} {\bibfnamefont {J.}~\bibnamefont
  {Castillo}}, \bibinfo {author} {\bibfnamefont {I.}~\bibnamefont {Vega}}, \
  and\ \bibinfo {author} {\bibfnamefont {B.}~\bibnamefont {Wardell}},\ }\href
  {\doibase 10.1103/PhysRevD.98.024024} {\bibfield  {journal} {\bibinfo
  {journal} {Phys. Rev. D}\ }\textbf {\bibinfo {volume} {98}},\ \bibinfo
  {pages} {024024} (\bibinfo {year} {2018})}\BibitemShut {NoStop}%
\bibitem [{\citenamefont {Pound}\ \emph {et~al.}(2005)\citenamefont {Pound},
  \citenamefont {Poisson},\ and\ \citenamefont {Nickel}}]{Poun05}%
  \BibitemOpen
  \bibfield  {author} {\bibinfo {author} {\bibfnamefont {A.}~\bibnamefont
  {Pound}}, \bibinfo {author} {\bibfnamefont {E.}~\bibnamefont {Poisson}}, \
  and\ \bibinfo {author} {\bibfnamefont {B.~G.}\ \bibnamefont {Nickel}},\
  }\href {\doibase 10.1103/PhysRevD.72.124001} {\bibfield  {journal} {\bibinfo
  {journal} {Phys. Rev. D}\ }\textbf {\bibinfo {volume} {72}},\ \bibinfo
  {pages} {124001} (\bibinfo {year} {2005})}\BibitemShut {NoStop}%
\bibitem [{\citenamefont {Bini}\ \emph {et~al.}(2007)\citenamefont {Bini},
  \citenamefont {Geralico},\ and\ \citenamefont {Ruffini}}]{Bini07}%
  \BibitemOpen
  \bibfield  {author} {\bibinfo {author} {\bibfnamefont {D.}~\bibnamefont
  {Bini}}, \bibinfo {author} {\bibfnamefont {A.}~\bibnamefont {Geralico}}, \
  and\ \bibinfo {author} {\bibfnamefont {R.}~\bibnamefont {Ruffini}},\ }\href
  {\doibase 10.1103/PhysRevD.75.044012} {\bibfield  {journal} {\bibinfo
  {journal} {Phys. Rev. D}\ }\textbf {\bibinfo {volume} {75}},\ \bibinfo
  {pages} {044012} (\bibinfo {year} {2007})}\BibitemShut {NoStop}%
\bibitem [{\citenamefont {{Martel}}\ and\ \citenamefont
  {{Poisson}}(2005)}]{MartPois05}%
  \BibitemOpen
  \bibfield  {author} {\bibinfo {author} {\bibfnamefont {K.}~\bibnamefont
  {{Martel}}}\ and\ \bibinfo {author} {\bibfnamefont {E.}~\bibnamefont
  {{Poisson}}},\ }\href {\doibase 10.1103/PhysRevD.71.104003} {\bibfield
  {journal} {\bibinfo  {journal} {Phys. Rev. D}\ }\textbf {\bibinfo {volume}
  {71}},\ \bibinfo {pages} {104003} (\bibinfo {year} {2005})}\BibitemShut
  {NoStop}%
\bibitem [{\citenamefont {Pugliese}\ \emph {et~al.}(2011)\citenamefont
  {Pugliese}, \citenamefont {Quevedo},\ and\ \citenamefont
  {Ruffini}}]{PuglQuevRuff11}%
  \BibitemOpen
  \bibfield  {author} {\bibinfo {author} {\bibfnamefont {D.}~\bibnamefont
  {Pugliese}}, \bibinfo {author} {\bibfnamefont {H.}~\bibnamefont {Quevedo}}, \
  and\ \bibinfo {author} {\bibfnamefont {R.}~\bibnamefont {Ruffini}},\ }\href
  {\doibase 10.1103/PhysRevD.83.104052} {\bibfield  {journal} {\bibinfo
  {journal} {Phys. Rev. D}\ }\textbf {\bibinfo {volume} {83}},\ \bibinfo
  {pages} {104052} (\bibinfo {year} {2011})}\BibitemShut {NoStop}%
\bibitem [{\citenamefont {Chandrasekhar}(1992{\natexlab{a}})}]{Chan92}%
  \BibitemOpen
  \bibfield  {author} {\bibinfo {author} {\bibfnamefont {S.}~\bibnamefont
  {Chandrasekhar}},\ }\enquote {\bibinfo {title} {The mathematical theory of
  black holes},}\ \ (\bibinfo  {publisher} {Oxford University Press, Oxford,
  United Kingdom},\ \bibinfo {year} {1992})\ Chap.~\bibinfo {chapter} {5}, pp.\
  \bibinfo {pages} {215--224}\BibitemShut {NoStop}%
\bibitem [{\citenamefont {Davis}\ \emph {et~al.}(1972)\citenamefont {Davis},
  \citenamefont {Ruffini}, \citenamefont {Tiomno},\ and\ \citenamefont
  {Zerilli}}]{DaviRuffTiom72}%
  \BibitemOpen
  \bibfield  {author} {\bibinfo {author} {\bibfnamefont {M.}~\bibnamefont
  {Davis}}, \bibinfo {author} {\bibfnamefont {R.}~\bibnamefont {Ruffini}},
  \bibinfo {author} {\bibfnamefont {J.}~\bibnamefont {Tiomno}}, \ and\ \bibinfo
  {author} {\bibfnamefont {F.}~\bibnamefont {Zerilli}},\ }\href {\doibase
  10.1103/PhysRevLett.28.1352} {\bibfield  {journal} {\bibinfo  {journal}
  {Phys. Rev. Lett.}\ }\textbf {\bibinfo {volume} {28}},\ \bibinfo {pages}
  {1352} (\bibinfo {year} {1972})}\BibitemShut {NoStop}%
\bibitem [{\citenamefont {Warburton}\ \emph {et~al.}(2017)\citenamefont
  {Warburton}, \citenamefont {Osburn},\ and\ \citenamefont
  {Evans}}]{WarbOsbuEvan17}%
  \BibitemOpen
  \bibfield  {author} {\bibinfo {author} {\bibfnamefont {N.}~\bibnamefont
  {Warburton}}, \bibinfo {author} {\bibfnamefont {T.}~\bibnamefont {Osburn}}, \
  and\ \bibinfo {author} {\bibfnamefont {C.~R.}\ \bibnamefont {Evans}},\ }\href
  {\doibase 10.1103/PhysRevD.96.084057} {\bibfield  {journal} {\bibinfo
  {journal} {Phys. Rev. D}\ }\textbf {\bibinfo {volume} {96}},\ \bibinfo
  {pages} {084057} (\bibinfo {year} {2017})}\BibitemShut {NoStop}%
\bibitem [{\citenamefont {{Hopman}}\ and\ \citenamefont
  {{Alexander}}(2005)}]{Hopm05}%
  \BibitemOpen
  \bibfield  {author} {\bibinfo {author} {\bibfnamefont {C.}~\bibnamefont
  {{Hopman}}}\ and\ \bibinfo {author} {\bibfnamefont {T.}~\bibnamefont
  {{Alexander}}},\ }\href {\doibase 10.1086/431475} {\bibfield  {journal}
  {\bibinfo  {journal} {Astrophys. J.}\ }\textbf {\bibinfo {volume} {629}},\
  \bibinfo {pages} {362} (\bibinfo {year} {2005})}\BibitemShut {NoStop}%
\bibitem [{\citenamefont {Zimmerman}\ \emph {et~al.}(2013)\citenamefont
  {Zimmerman}, \citenamefont {Vega}, \citenamefont {Poisson},\ and\
  \citenamefont {Haas}}]{ZimmETC13}%
  \BibitemOpen
  \bibfield  {author} {\bibinfo {author} {\bibfnamefont {P.}~\bibnamefont
  {Zimmerman}}, \bibinfo {author} {\bibfnamefont {I.}~\bibnamefont {Vega}},
  \bibinfo {author} {\bibfnamefont {E.}~\bibnamefont {Poisson}}, \ and\
  \bibinfo {author} {\bibfnamefont {R.}~\bibnamefont {Haas}},\ }\href {\doibase
  10.1103/PhysRevD.87.041501} {\bibfield  {journal} {\bibinfo  {journal} {Phys.
  Rev. D}\ }\textbf {\bibinfo {volume} {87}},\ \bibinfo {pages} {041501}
  (\bibinfo {year} {2013})}\BibitemShut {NoStop}%
\bibitem [{\citenamefont {Sorce}\ and\ \citenamefont
  {Wald}(2017)}]{SorcWald17}%
  \BibitemOpen
  \bibfield  {author} {\bibinfo {author} {\bibfnamefont {J.}~\bibnamefont
  {Sorce}}\ and\ \bibinfo {author} {\bibfnamefont {R.~M.}\ \bibnamefont
  {Wald}},\ }\href {\doibase 10.1103/PhysRevD.96.104014} {\bibfield  {journal}
  {\bibinfo  {journal} {Phys. Rev. D}\ }\textbf {\bibinfo {volume} {96}},\
  \bibinfo {pages} {104014} (\bibinfo {year} {2017})}\BibitemShut {NoStop}%
\bibitem [{\citenamefont {Chandrasekhar}(1992{\natexlab{b}})}]{Chan92b}%
  \BibitemOpen
  \bibfield  {author} {\bibinfo {author} {\bibfnamefont {S.}~\bibnamefont
  {Chandrasekhar}},\ }\enquote {\bibinfo {title} {The mathematical theory of
  black holes},}\ \ (\bibinfo  {publisher} {Oxford University Press, Oxford,
  United Kingdom},\ \bibinfo {year} {1992})\ Chap.~\bibinfo {chapter} {11},
  pp.\ \bibinfo {pages} {573--580}\BibitemShut {NoStop}%
\bibitem [{\citenamefont {Zhang}(2016)}]{Zhan16}%
  \BibitemOpen
  \bibfield  {author} {\bibinfo {author} {\bibfnamefont {B.}~\bibnamefont
  {Zhang}},\ }\href {\doibase 10.3847/2041-8205/827/2/L31} {\bibfield
  {journal} {\bibinfo  {journal} {Astrophys. J. Lett.}\ }\textbf {\bibinfo
  {volume} {827}},\ \bibinfo {pages} {L31} (\bibinfo {year}
  {2016})}\BibitemShut {NoStop}%
\bibitem [{\citenamefont {Ishibashi}\ and\ \citenamefont
  {Kodama}(2011)}]{IshiKoda11}%
  \BibitemOpen
  \bibfield  {author} {\bibinfo {author} {\bibfnamefont {A.}~\bibnamefont
  {Ishibashi}}\ and\ \bibinfo {author} {\bibfnamefont {H.}~\bibnamefont
  {Kodama}},\ }\href {https://ci.nii.ac.jp/naid/210000128959/} {\bibfield
  {journal} {\bibinfo  {journal} {Prog. of Theor. Phys. Suppl.}\ }\textbf
  {\bibinfo {volume} {189}},\ \bibinfo {pages} {165} (\bibinfo {year}
  {2011})}\BibitemShut {NoStop}%
\end{thebibliography}%

\end{document}